\documentclass[lettersize,journal]{IEEEtran}
\usepackage{amsmath,amsfonts}
\usepackage{array}
\usepackage[caption=false,font=normalsize,labelfont=sf,textfont=sf]{subfig}
\usepackage{textcomp}
\usepackage{stfloats}
\usepackage{url}
\usepackage{verbatim}
\usepackage{graphicx}
\usepackage{cite}

\usepackage{soul,listings,xcolor}

\usepackage{hyperref}

\usepackage{graphicx}

\lstnewenvironment{teX}[1][]
{\lstset{language=[LaTeX]TeX}\lstset{escapeinside={(*@}{@*)},
		numbers=left,numberstyle=\scriptsize,stepnumber=1,numbersep=5pt,
		breaklines=true,
		framesep=5pt,
		basicstyle=\scriptsize\ttfamily,
		showstringspaces=false,
		keywordstyle=\itshape\color{blue},
		stringstyle=\color{maroon},
		commentstyle=\color{black},
		rulecolor=\color{black},
		xleftmargin=0pt,
		xrightmargin=0pt,
		aboveskip=\medskipamount,
		belowskip=\medskipamount,
		backgroundcolor=\color{white}, #1
}}
{}

\usepackage{algorithm}
\usepackage{algorithmicx}
\usepackage{algpseudocode}
\usepackage{amsmath}

\usepackage{booktabs}

\usepackage{multirow}

\usepackage{threeparttable}

\usepackage[misc]{ifsym}

\usepackage {textcase}

\usepackage{makecell}

\usepackage{bbding}

\begin{document}

\title{\textsc{Instiller}: Towards Efficient and Realistic RTL Fuzzing}

\author{Gen Zhang, Pengfei Wang\textsuperscript{\Letter}, Tai Yue, Danjun Liu, Yubei Guo and Kai Lu\textsuperscript{\Letter} \\National University of Defense Technology}



\maketitle

\begin{abstract}
Bugs exist in hardware, such as CPU. Unlike software bugs, these hardware bugs need to be detected before deployment. Previous fuzzing work in CPU bug detection has several disadvantages, e.g., the length of RTL input instructions keeps growing, and longer inputs are ineffective for fuzzing.

In this paper, we propose \textsc{Instiller} (\textbf{Inst}ruction Di\textbf{stiller}), an RTL fuzzer based on ant colony optimization (ACO). First, to keep the input instruction length short and efficient in fuzzing, it distills input instructions with a variant of ACO (VACO). Next, related work cannot simulate realistic interruptions well in fuzzing, and \textsc{Instiller} solves the problem of inserting interruptions and exceptions in generating the inputs. Third, to further improve the fuzzing performance of \textsc{Instiller}, we propose hardware-based seed selection and mutation strategies.

We implement a prototype and conduct extensive experiments against state-of-the-art fuzzing work in real-world target CPU cores. In experiments, \textsc{Instiller} has 29.4\% more coverage than DiFuzzRTL. In addition, 17.0\% more mismatches are detected by \textsc{Instiller}. With the VACO algorithm, \textsc{Instiller} generates 79.3\% shorter input instructions than DiFuzzRTL, demonstrating its effectiveness in distilling the input instructions. In addition, the distillation leads to a 6.7\% increase in execution speed on average.
\end{abstract}


\section{Introduction} \label{sec_intro}
CPU bugs are notorious. Besides the well-known Meltdown and Spectre \cite{meltdown}, numerous bugs are reported, including the Pentium FDIV bug \cite{fdiv}, Broadwell MCE bug \cite{mce}, and Ryzen segfault bug \cite{ryzen}. All of them can cost the manufacturers millions or billions of dollars in mitigating and repairing the bugs.

Before CPU deployment, the circuits and RTL (register transition level) designs should be thoroughly verified. In software deployment, bugs can be avoided with timely patches. However, in the development of CPU, once deployed, it is nearly impracticable to remove the impact of hardware vulnerabilities. For example, the mitigation of Meltdown and Spectre only focuses on part of the mainstream products, due to the challenging balance among the mitigation itself, performance impact, and implementation complexity \cite{mitigation}.

Previous work has made some attempts to detect CPU bugs \cite{moundanos1998abstraction, kantrowitz1996m}, both in static and dynamic techniques. Among them, verifying CPU with fuzz testing \cite{miller1990empirical} is one of the most promising approaches \cite{rfuzz, difuzzrtl}. However, there are still several drawbacks to these techniques, and they extend to the following challenges.


\textbf{Challenge 1: growing input length. }The basic structure of the input instructions of the CPU consists of an instruction sequence. Interruptions and exceptions can be inserted to simulate the real-world execution of the CPU. Starting from a simple instruction, as the fuzzing process goes on, the length of input tends to increase. The speed of fuzz testing is the key to its success \cite{zeror, fullspeed, instrcr, wang2021riff, wang2022odin, li2023accelerating}. Longer input instructions are catastrophic to the execution speed of fuzzing since longer inputs will spend more CPU cycles. More importantly, according to our analysis in the evaluation in Section \ref{sec_length}, coverage does not increase proportionally to input length. Therefore, we should come up with solutions to shorten the input instructions and improve the fuzzing efficiency.

\textbf{Challenge 2: realistic interruption and exception handling. } Interruptions and exceptions are common in the execution of the CPU. Simulating them in testing CPU can cover the corner cases of CPU verification. Previous fuzzing work mentioned considering interruptions in the design \cite{difuzzrtl}, but the approach is relatively simple. Exceptions are not simulated, and multiple interruptions and exceptions and their priorities are also not included. Missing these situations cannot simulate the real-world CPU execution and cannot cover all the CPU states.

\textbf{Challenge 3: fuzzing techniques related to hardware. }Fuzzing techniques are initially designed for testing software. Many customized approaches in fuzzing, such as program transformation \cite{tfuzz} and process tracing \cite{schumilo2017kafl}, are also tailored for software programs. Especially in the critical steps of fuzzing, such as seed selection and mutation, previous fuzzing work did not combine fuzzing with hardware features well. For example, \cite{difuzzrtl} did not consider seed selection when fuzzing the CPU. Not using hardware-related techniques cannot improve the fuzzing performance in testing CPU RTL.

To address the above challenges, we propose the following techniques.

\textbf{Input instruction distillation based on a variant of ant colony optimization. }To save the CPU cycles and improve fuzzing performance, we need to distill the inputs and shorten the input instruction length. The basic idea of input instruction distillation is to construct a subset of the original input set, which is shorter in length and can maintain the original coverage. Ant colony optimization (ACO) is one of the latest techniques for approximate optimization \cite{blum2005ant}. The algorithm simulates the routine of an ant colony to search for the shortest path to the target city. We use the idea of ACO to distill input instructions. We model the length of input instructions as the number of ants and the RTL circuits as cities. The algorithm can output the best input instruction and length for the current status, which finishes the task of input instruction distillation. Moreover, we make some changes to the classic ACO and propose a variant of ACO (VACO) to fit the RTL fuzzing scenario.

\textbf{Simulating realistic interruption and exception handling. }First, we include exceptions in fuzzing the CPU, which is not proposed in previous work \cite{difuzzrtl}. Next, we integrate more than one interruption and exception to test the CPU, aiming to simulate the real-world execution scenario of the CPU more comprehensively. Besides, we consider the priorities of different interruptions and exceptions, which can thoroughly fuzz the CPU. The above techniques can better simulate real-world interruption and exception handling than previous work.

\textbf{Hardware-related seed selection and mutation. }We propose new seed selection and mutation strategies in fuzzing the CPU. In seed selection, not only basic fuzzing heuristics are taken into consideration, but also hardware heuristics, e.g., special instructions and registers. For mutation, we propose strategies closely related to hardware, such as insertion or deletion based on the input instruction length. The seed selection and mutation strategies combine fuzzing with hardware characteristics, and they overcome the drawbacks of previous tools.

We implement a prototype \textsc{Instiller} (\textbf{Inst}ruction Di\textbf{stiller}) and conduct extensive evaluation against state-of-the-art fuzzing work. In general, the results show the effectiveness of our proposed techniques. Our tool increases coverage by 29.4\%. For input instruction distillation, the length of \textsc{Instiller} is 79.3\% shorter than DiFuzzRTL. For vulnerability discovery, \textsc{Instiller} finds 17.0\% more mismatches in the targets. In addition, the input instruction distillation leads to a 6.7\% increase in execution speed on average.

In conclusion, we make the following contributions to this paper:

\begin{itemize}
	\item We propose an input instruction distillation technique, which is based on a variant of ant colony optimization. The distillation can make the inputs shorter and more effective.
	\item We enable our fuzzer to handle multiple interruptions and exceptions. The priorities of them are also considered. These techniques can simulate realistic interruption and exception handling well.
	\item We propose hardware-based seed selection and mutation strategies. We use hardware-related heuristics and mutation operations to improve fuzzing performance in the situation of RTL fuzzing.
	\item We implement a prototype named \textsc{Instiller} and conduct extensive experiments. The results show that our tool outperforms previous work and demonstrate the effectiveness of our proposed approaches.
\end{itemize}

\section{Background and Motivation}

\subsection{CPU Design, Interruptions, and Exceptions} \label{sec_2_A}

\textbf{CPU design and verification. }ISA (instruction set architecture) is the basis of designing a CPU. Different types of ISA can be implemented on different processors, e.g., Intel (X86) and M1 (ARM). RTL is a real hardware design based on the specific ISA. RTL can be described in hardware description languages (HDL) such as Verilog. Most importantly, CPU RTL should be completely tested before deployment. Not like software, the bugs and vulnerabilities in hardware cannot be easily mitigated with patches. Dynamic techniques are more common in testing RTL \cite{difuzzrtl}, which include testing with ISA and RTL simulation. The idea of differential testing in \textsc{Instiller} is to compare the results of ISA and RTL simulation and detect hardware bugs.

\textbf{Input instructions, interruptions, and exceptions. } Before execution, specific instructions should be loaded into the CPU. In this paper, the inputs of CPU RTL are formed with different instructions. By interpreting every instruction, the CPU finishes executing the program. Besides normal executions, interruptions and exceptions will raise, e.g., IO interruptions and illegal access exceptions. When testing RTL, interruptions and exceptions should be simulated to thoroughly cover the corner situations. Moreover, multiple interruptions and exceptions and their priorities are common in CPU execution. For example, another high-priority interruption occurs when handling a low-priority one. Previous fuzzing work \cite{difuzzrtl, rfuzz} failed to handle multiple interruptions or exceptions and their priorities. We consider these situations in our paper.

\vspace{-0.4cm}

\begin{figure}[htbp]
	\setlength{\abovecaptionskip}{-0.1cm}
	\centering
	\includegraphics[width=0.7\columnwidth]{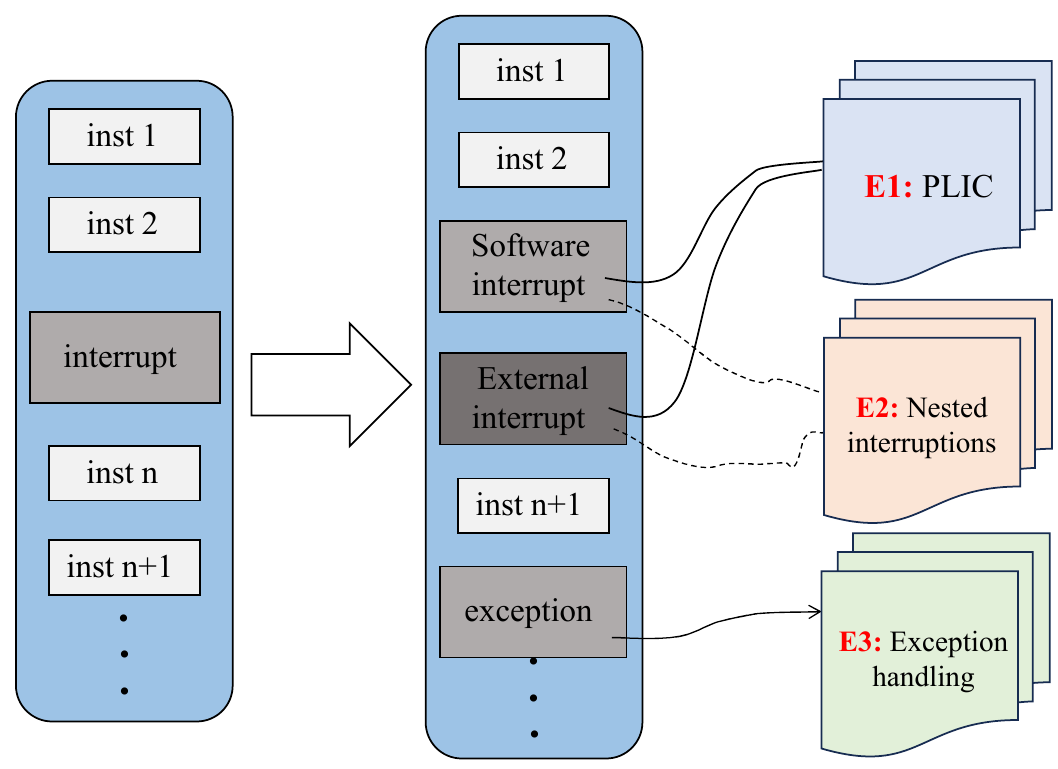}
	\caption{{Example of the effectiveness of multi-interruption and exception hardware fuzzing.}}
	\label{fig_multi_intr_cov}
\end{figure}

\vspace{-0.4cm}

Figure \ref{fig_multi_intr_cov} is a motivating example showing how a single-interruption no-exception strategy limits the effectiveness of fuzzing. This is a simplified input sequence containing instructions, two interruptions with different privilege levels, and an exception. There are mainly three aspects where the single-interruption no-exception strategy limits the effectiveness of fuzzing, i.e., coverage. First, platform level interrupt controller (PLIC) is a subsystem that can control the arbitration and distribution of multiple interrupts \cite{plic}. The code in this subsystem can only be triggered when there are multiple interruptions in the input sequence. Thus, single-interruption fuzzers fail to reach this coverage, which is denoted as \textit{E1} in Figure \ref{fig_multi_intr_cov}.

Moreover, the single-interruption no-exception strategy also hinders the fuzzers from covering nested interruption scenarios, which is denoted as \textit{E2}. Handling nested interruptions is enabled in certain IP cores \cite{volumeII}. Single interruption can never reach these scenarios. Therefore, enabling multiple and nested interruptions in CPU fuzzing can reach more coverage.

Third, exception handling is an important subsystem in CPU implementation \cite{waterman2014risc}, and there are at least 16 entries for standard exception handling. No-exception fuzzing strategy fails to cover this exception-handling code. Our proposed fuzzing strategy with exceptions in the input sequence can solve this problem, which is denoted as \textit{E3} in Figure \ref{fig_multi_intr_cov}.

\begin{figure}[htbp]
	\setlength{\abovecaptionskip}{-0.1cm}
	\centering
	\includegraphics[width=0.8\columnwidth]{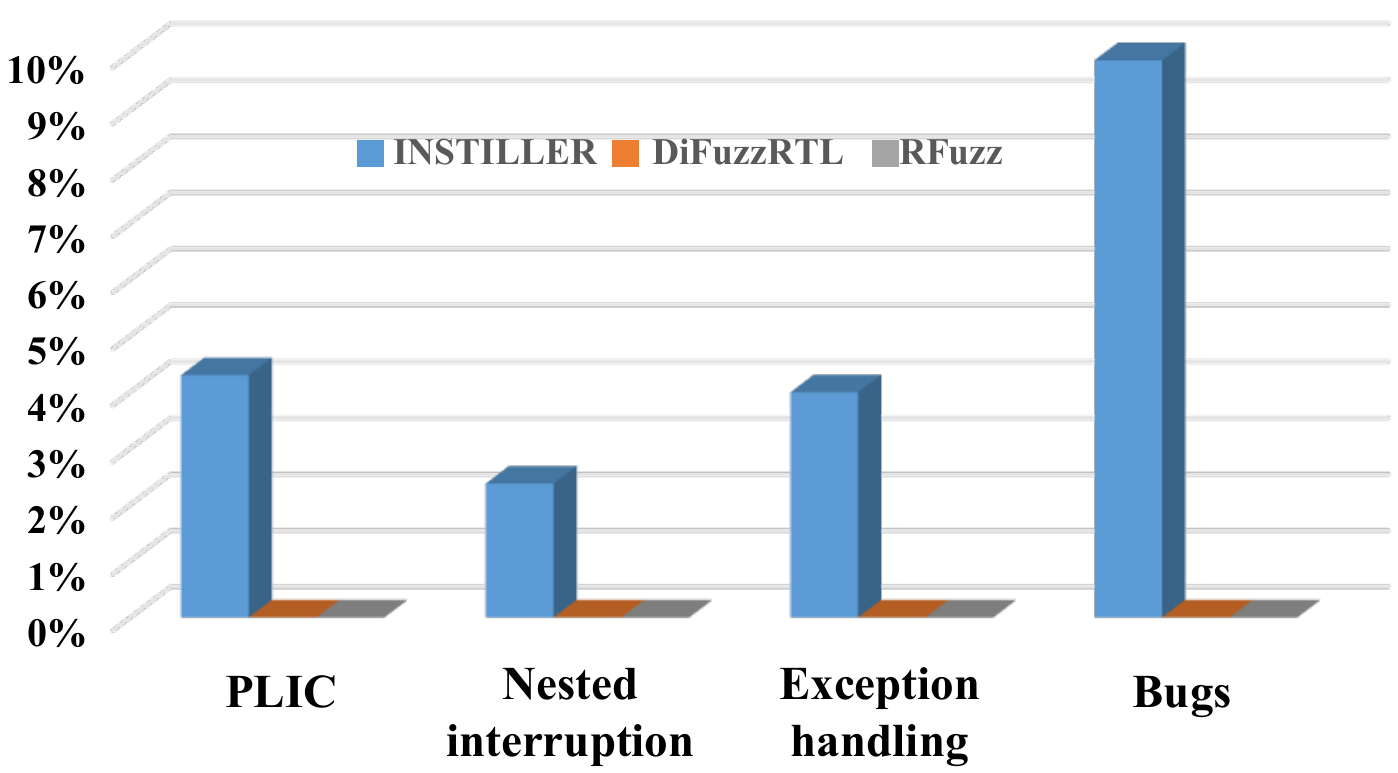}
	\caption{{Percentages of HDL line coverage and bugs of \textsc{Instiller}, DiFuzzRTL, and RFuzz related to multiple interruptions and exceptions.}}
	\label{fig_multi_intr_cov_back}
\end{figure}

\vspace{-0.4cm}

Moreover, a preliminary experiment is conducted to reveal the insight of multiple interruptions and exceptions. We compare \textsc{Instiller} with state-of-the-art DiFuzzRTL and RFuzz to collect the coverage increase and bug detection related to interruptions and exceptions in Figure \ref{fig_multi_intr_cov_back}. DiFuzzRTL and RFuzz are not designed with multiple interruptions and exceptions. Therefore, their related line coverage and detected bugs are zeros. The percentages of HDL lines covered by \textsc{Instiller} are 4.3\%, 2.4\%, and 4.0\% of the overall coverage, in PLIC, nested interruptions, and exception handling, respectively. In total, HDL line coverage related to multiple interruptions and exceptions makes up for over 10\% of the coverage by \textsc{Instiller}. Besides, bugs triggered by multiple interruptions and exceptions are about 10\% of all the bugs. The results in Figure \ref{fig_multi_intr_cov_back} are \textbf{quantitative examples} of the motivation in Figure \ref{fig_multi_intr_cov}.

Therefore, we believe enabling multiple interruptions and exceptions with their priorities is innovative compared with other fuzzers in two aspects. First, the fuzzing strategy in \cite{difuzzrtl, rfuzz} is immature without multiple interruptions and exceptions with their priorities, and they deviate significantly from the actual CPU execution scenario. This is the key difference between our proposed ``REALISTIC" strategy and theirs. Second, we have invested sufficient effort in code implementation in this part. For instance, the insertion of multiple interruptions with privilege levels requires us to deal with the \textit{mcause} register and other issues. Therefore, our strategy is creative compared with previous work.

\subsection{Fuzzing and Differential Testing}


\begin{figure*}[htbp]
	\setlength{\abovecaptionskip}{-1cm}
	\centering
	\includegraphics[width=0.99\textwidth]{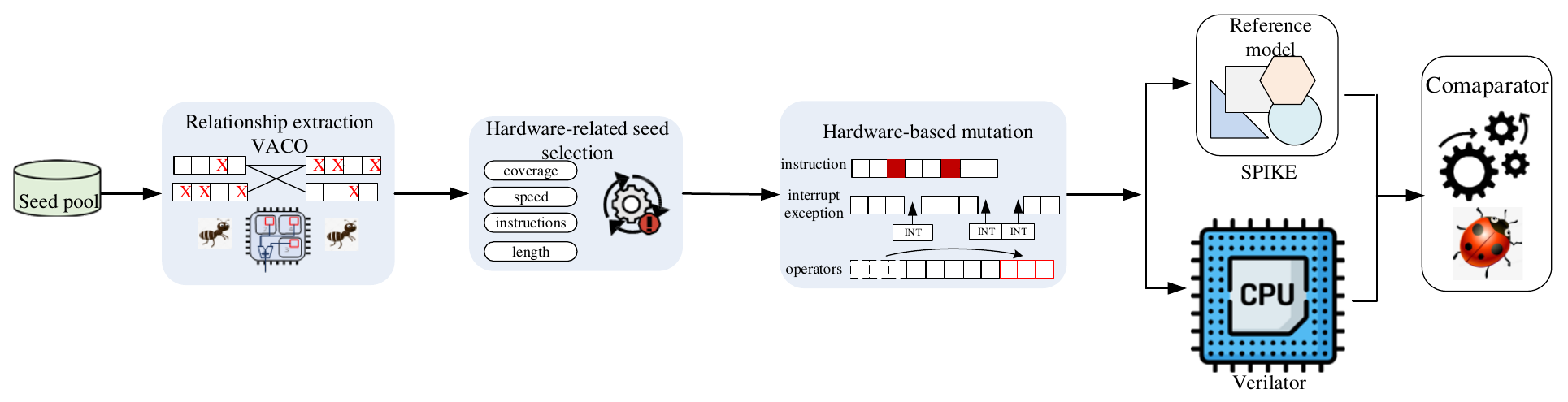}
	\caption{{Overview of the basic procedures in \textsc{Instiller}, including VACO, seed selection and mutation, and verilator.}}
	\label{fig_overview}
\end{figure*}

Fuzzing or fuzz testing \cite{miller1990empirical} is one of the most successful software testing techniques. Coverage-guided grey-box fuzzing (CGF) is a variant of fuzzing, and it is famous for its great balance between effectiveness and efficiency \cite{afl}. The basic steps of fuzzing include:

\begin{itemize}
	\item 1) Given initial seeds;
	\item 2) Select a seed and mutate the seed to generate input instructions;
	\item 3) Execute the target program with the instructions;
	\item 4) If there is new coverage, save the input instruction to the seed pool; If there is a program crash, report and save this crash;
	\item 5) Continuing to step 2).
\end{itemize}     

Differential testing compares the results of two or more systems, and different results indicate bugs and vulnerabilities \cite{mckeeman1998differential}. Based on it, differential fuzz testing is proposed to solve more complex bug-discovery problems \cite{difuzzrtl}.

In CPU verification, comparing the results of RTL executions with that of a golden model (ISA) is an effective testing technique \cite{difuzzrtl, kande2022thehuzz}. Therefore, our work combines the idea of CGF and differential testing to boost the CPU verification process.

\subsection{Ant Colony Optimization}

Ant colony optimization was first introduced in the early 1990’s \cite{blum2005ant}. It is one of the latest techniques for approximate optimization. ACO is used to solve combinatorial optimization problems such as the traveling salesman problem (TSP), which is an optimization algorithm simulating ant foraging behavior.



The basic procedure of ACO can be concluded as:
\begin{itemize}
	\item 1) Initialize the parameters;
	\item 2) Calculate the probability for every city and every ant;
	\item 3) Select the best next city to walk for every ant;
	\item 4) Update the pheromone table after the ants finish walking;
	\item 5) Continue to step 2) before termination.
\end{itemize}

In this paper, we use the idea of ACO to distill input instructions. However, classic ACO does not fit RTL testing. The number of ants in ACO is constant, and it is variable in our model. Moreover, in classic ACO, the algorithm chooses the next best city for an ant. In our work, executing an input instruction will cover multiple circuits (cities), and it is choosing the next best cities. Therefore, we propose a variant of classic ACO to handle the above problems.

\section{Design}

\subsection{Overview}

Figure \ref{fig_overview} is the overview of \textsc{Instiller}. There are mainly three newly-designed infrastructures, including the VACO algorithm, interruption and exception simulation, and seed selection with mutation. First, VACO is capable of distilling the input instructions in RTL fuzzing, which can keep the input short and effective. Next, realistic interruption and exception handling is simulated by our simulation process. Through this kind of simulation, our fuzzing process is closer to the real-world execution of the CPU. In addition, the seed selection and mutation strategies integrate hardware-related features into fuzzing and improve the fuzzing performance.

\begin{algorithm}
	\caption{Overview of the procedures of \textsc{Instiller}.}
	\label{algo_overview}
	\scriptsize
	\begin{algorithmic}[1]
		\Require Initial seeds $S$
		\While{$t < TIME\_OUT$}
		\If{$start\_distill == True$}
		\State $re = relation\_extract()$
		\State $len = VACO(re)$
		\EndIf
		\State $s = seed\_selection(seed)$
		\State $s' = mutation(s, len)$
		\State $input = interrupt\_exception(s')$
		\State $O_I=ISA\_sim(input)$
		\State $O_R=RTL\_sim(input)$
		\State $Cross\_check(O_I, O_R)$
		\EndWhile
		\Ensure Bug reports
	\end{algorithmic}
\end{algorithm}

\begin{figure}[htbp]
	\setlength{\abovecaptionskip}{-0.5cm}
	\centering
	\includegraphics[width=\columnwidth]{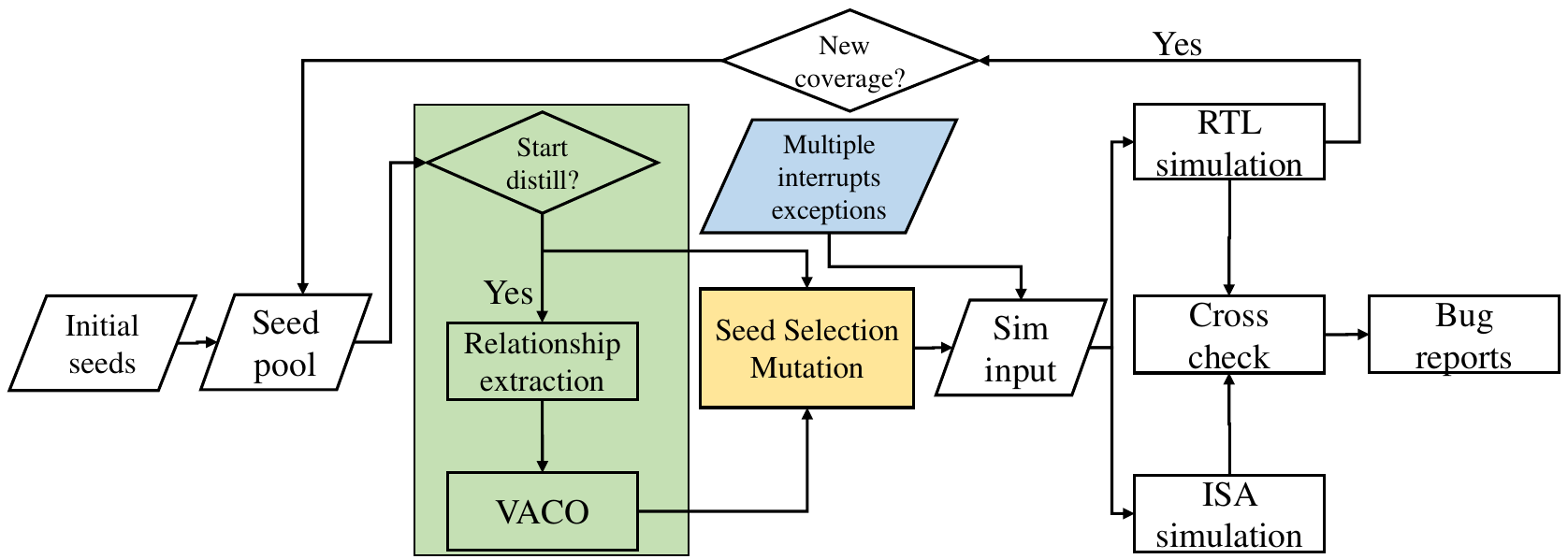}
	\caption{{Fuzzing procedures of \textsc{Instiller}, where the colored parts are newly proposed mechanism compared with traditional fuzzing.}}
	\label{fig_fuzz_overview}
\end{figure}

The detailed execution process of \textsc{Instiller} is shown in Algorithm \ref{algo_overview}. Given initial seeds, the fuzzing process is started. Depending on the current coverage status, \textsc{Instiller} decides whether the input instruction distillation should be started. Distillation includes relationship extraction and the VACO algorithm. The output of distillation is the most effective input and its length for the current fuzzing status. After seed selection and mutation, the input instructions are inserted with multiple interruptions and exceptions, which are ready for execution. ISA simulation and RTL simulation will be executed, and their results are cross-checked to output bug reports. In general, Figure \ref{fig_fuzz_overview} shows the fuzzing procedure of \textsc{Instiller}, and the colored parts are the modification to the basic fuzzing process.

\vspace{-0.4cm}

\subsection{Input instruction Distillation Based on VACO}

In state-of-the-art RTL fuzzing work \cite{difuzzrtl}, the length of the input instruction keeps increasing as the fuzzing continues. Long input instructions slow down the fuzzing process and are unfriendly for fuzzing. Therefore, we propose input instruction distillation to keep the inputs short and effective. The distillation includes relationship extraction and the VACO algorithm.

\begin{figure}[htbp]
	\setlength{\abovecaptionskip}{-0.5cm}
	\centering
	\includegraphics[width=\columnwidth]{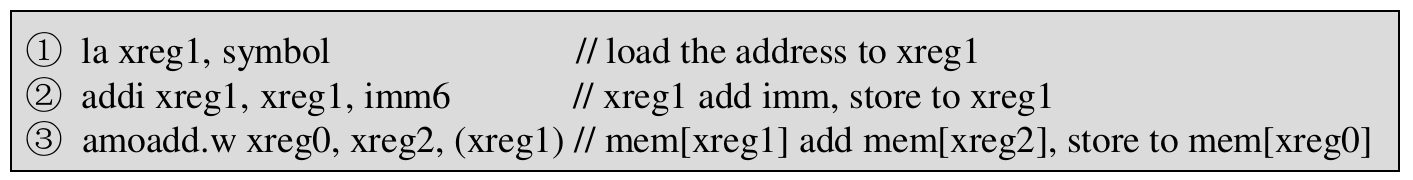}
	\caption{{Example of the relationships between instructions, where the three instructions all operate on the same register.}}
	\label{fig_relation}
\end{figure}

\textbf{Relationship extraction. }In the execution of the CPU, the complex operations are finished by some related instructions and they should be treated as a group. For example, in Figure \ref{fig_relation}, the three instructions complete an $ADD$ operation. Therefore, it is rational to extract the relationships between instructions. According to our preliminary study \footnote{This study is conducted by investigating the RISC-V Instruction Set Manual Volume I and II \cite{waterman2014risc, volumeII}.}, the relationships between RTL input instructions can be divided into software relationships and hardware relationships. If there are relationships between instructions, we collect them to form instruction groups. These groups are then used in the VACO algorithm.

\begin{algorithm}
	\caption{The relationship extraction algorithm.}
	\label{algo_relation}
	\scriptsize
	\begin{algorithmic}[1]
		\Require Software inputs $W_s$, hardware inputs $W_h$, coverage $cov$
		\State $W=sort(W_s, cov)$
		\State $W_D = \emptyset $ /* instruction groups */
		\For{$i \to W$}
		\For {$j \to W$}
		\If {$register_i == register_j$}
		\State $W_D = W_D + (i, j)$
		\EndIf
		\If {$target(i) == j$}
		\State $W_D = W_D + (i, j)$
		\EndIf
		\EndFor
		\EndFor
		\For {$i \to W_h$}
		\If {$i \Rightarrow clock, iterrupt, privilege, register $}
		\State $W_D = W_D + i$
		\EndIf
		\EndFor
		\Ensure Instruction groups $W_D$
	\end{algorithmic}
\end{algorithm}

Software relationships include data-flow and control-flow relationships. First, we sort all the executed input instructions with coverage. Then, beginning from the input instructions with the most coverage, the instructions are traversed. If two instructions share the same registers, there is a data-flow relationship between them. If one input is the jump target of another, there is a control-flow relationship between them. Instructions that have software relationships are collected to form groups.

Hardware relationships include clock cycles, interruptions, privilege levels, and special registers. For example, if inserting an interruption after an instruction can change the privilege level (priority), we consider there is a hardware relationship. Another example is the non-aligned load and store addresses are exceptional. When the hart time comparator (a memory mapping register named $\mathtt{mtimecmp}$) is larger than the real-time counter $\mathtt{mtime}$, the clock interruption will be triggered. These instructions together with the hardware events are collected to form groups.

Algorithm \ref{algo_relation} shows the process of the relationship extraction procedure. This approach outputs the instruction groups to VACO.

\textbf{The VACO algorithm. }As we discussed above, by simulating the behaviors of the ant colony, ACO is a classic optimization algorithm to find the shortest path between cities. It is an iterative algorithm, and it outputs the best solution when the iteration is done. To adopt ACO in our scenario, we need to solve three problems: 1) When to start the algorithm? 2) How to model the factors in fuzzing into the algorithm? 3) When to stop the process?

1) The algorithm is invoked when there is a continuous average coverage decrease. Average coverage is coverage divided by the input instruction length. When this indicator decreases, the input instruction length is too long, and the input is not effective in finding new coverage.

2) We model the length of the input instructions as the number of ants, and the RTL circuits as the cities in ACO. The scale of RTL \footnote{In this paper, we use the coverage definition of DiFuzzRTL. Therefore, the scale of RTL means the number of control registers \cite{difuzzrtl}.} (number of cities) is $\mathtt{n}$, and the current length of input instruction (number of ants) is $\mathtt{m}$. In each iteration, for every ant $\mathtt{i}$ and every path $\mathtt{j}$, we calculate the pheromone table as:

\begin{equation}
	\label{equ_pheromone}
	pher_j = (1-\rho)*pher_j+\sum_{i=1}^m\Delta pher_j^i
\end{equation}

In Equation \ref{equ_pheromone}, $\rho$ means the evaporation rate of pheromones, which is a tunable parameter. The definition of $\Delta pher_j^i$ is:

\begin{equation}
	\label{equ_pher_delta}
	\Delta pher_j^i = \left\{
	\begin{array}{ll}
		\frac{1}{length_i} & ant\;i\;traverses\;path\;j\\
		0 & otherwise\\
	\end{array}
	\right.
\end{equation}

Then, a probability table is calculated by:

\begin{equation}
	\label{equ_probability}
	p_j^i=\frac{pher_j*h_j}{\sum\limits_{k\,not\,traversed\,by\,i}^{n}pher_k*h_k}
\end{equation}

This equation uses coverage as heuristics ($h_j$), which is proportional to the coverage of path $\mathtt{j}$. This is where we use the instruction groups. Every instruction group stands for an ant. Executing the instructions in a group is the process of an ant walking through the cities. The best candidate for the next ant is selected with probability table $p_j^i$. Therefore, by appending the best group in this iteration to the current input instructions, the length of input will increase by one after each iteration of the algorithm. When the algorithm is terminated, the length of the input instructions is the most effective length for the current fuzzing status.

3) The algorithm will iterate until there is an average coverage increase compared with that before starting the algorithm.

\vspace{-0.2cm}
\begin{figure}[htbp]
	\setlength{\abovecaptionskip}{-0.2cm}
	\centering
	\includegraphics[width=0.6\columnwidth]{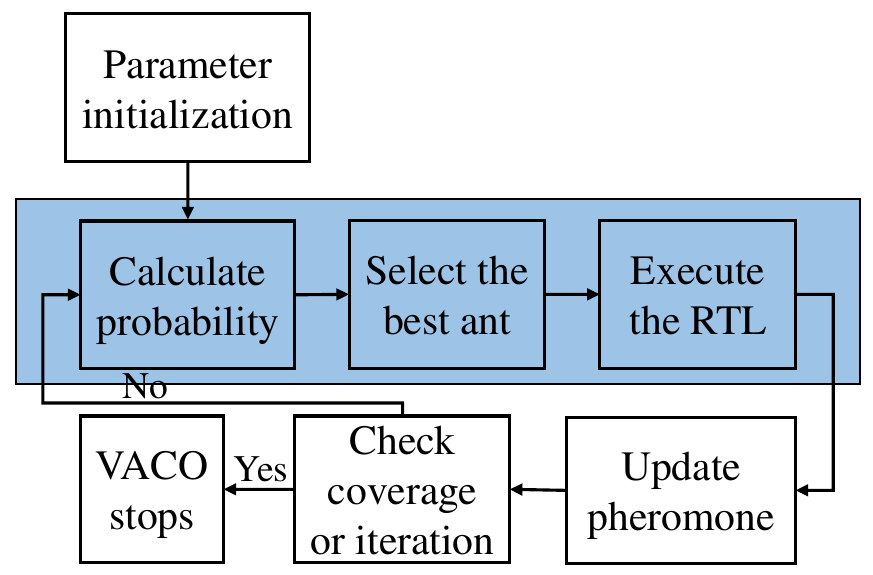}
	\caption{{Basic workflow of VACO, where the colored parts are newly designed compared with ACO.}}
	\label{fig_VACO}
\end{figure}
\vspace{-0.4cm}

As Figure \ref{fig_VACO} shows, the basic workflow is an iterative algorithm. Our proposed algorithm is a variant of the classic ACO. The colored parts in the figure are the major differences between VACO and ACO: 1) The number of ants in ACO is constant, while it is variable in VACO. When the algorithm is terminated, the generated length is the best input instruction length. 2) In VACO, the probability calculation is based on the coverage of the fuzzing process. However, the transition probability in ACO is related to the distance between cities. 3) The classic ACO does not contain the process of executing the ant in the RTL. Therefore, introducing the execution into VACO inevitably causes performance overhead. We integrate the RTL execution in VACO into the main fuzzing loop to reduce the performance overhead.

\begin{algorithm}
	\caption{The VACO algorithm.}
	\label{algo_vaco}
	\scriptsize
	\begin{algorithmic}[1]
		\Require Number of iterations $Max\_iter$, evaporation rate $\rho$, RTL scale $n$, input length $m$, group $W_D$
		\While{$iteration < Max\_iter$}
		\State $pher[:n] = [1...1]$
		\State $heuristics[:iteration] = coverage$
		\State $p = Equation\,\ref{equ_probability}$
		\State $id = roulette\_wheel\_selection(p)$ 
		\State $coverage' = exe(id)$ /* Executing this input instruction results in coverage' */
		\If {$average\;coverage\;increased$}
		\State $break$
		\EndIf
		\State $pher = (1 - \rho) * pher + \Sigma \Delta pher$
		\EndWhile
		\State $m = m + len(id)$
		\Ensure The most effective input instruction length $m$
	\end{algorithmic}
\end{algorithm}

In conclusion, Algorithm \ref{algo_vaco} shows the process of the VACO algorithm. By receiving the parameters, VACO iterates and finally outputs the best input instruction and length in the current fuzzing status. The length is shorter and the input instruction is more effective in finding new coverage. Therefore, the input distillation is finished. In Line 5, roulette wheel is a classic selection algorithm \cite{lipowski2012roulette}.

\subsection{Realistic Interruption and Exception Handling}

First, previous work has no exception inserted in the input instructions. We include exceptions in the RTL inputs to solve this problem. Next, in real-world applications, multiple interruptions and exceptions are common during CPU execution. We consider this situation in RTL testing and enable multiple interruptions and exceptions together with their respective handling strategies. In our design, multiple interruptions and exceptions are inserted in the input instructions for simulation. Every interruption and exception is analyzed to extract the relative information, e.g., the address of the interruption and the cause of the exception.


Figure \ref{fig_multi_intr_cov} shows the difference between multiple interruptions and exceptions and single ones. Several instructions $\mathtt{inst{\,}n}$ constitute a complete input. Interruptions and exceptions can be inserted between the instructions. The input on the right with multiple interruptions or exceptions is the one in \textsc{Instiller}. This input can better simulate real-world CPU execution.

Moreover, interruptions and exceptions have specific priorities. For example, in RISC-V, the priority of interruptions in machine mode is higher than that in supervisor mode \cite{waterman2014risc}. We consider different interruptions and exceptions with different priorities in fuzzing. In a situation where interruptions and exceptions with higher priority are triggered when a lower-priority one is being handled, more RTL states can be covered than testing without priorities.

In conclusion, in the process of fuzzing the RTL, we include both multiple interruptions and exceptions along with their priorities to test the target RTL more thoroughly.

\subsection{Seed Selection}

Previous work in RTL fuzzing ignores the importance of seed selection and hardware features in improving the performance of fuzzing. In seed selection, we focus on the normalized heuristics of input instructions to make decisions. Normalized heuristics means the heuristics score of an input divided by its length. The reason for using normalized heuristics is that one important perspective of our work is to distill seeds and generate shorter input instructions. A higher normalized heuristics score indicates the input instruction is more potential for fuzzing, and its length is relatively short at the same time.

In \textsc{Instiller}, we use the following heuristics to select seeds. 1) Basic heuristics. It includes coverage increase ($\mathtt{cov}$) and execution speed ($\mathtt{speed}$). These metrics of heuristics are commonly seen in fuzzing tools \cite{bohme2016coverage, yue2020ecofuzz, zhang2022mobfuzz}. We use these metrics to score seeds in this paper. Specifically, coverage is the most important metric in coverage-guided fuzzing. Therefore, it is used in our tool. Then, execution speed is also a crucial factor in fuzzing \cite{zeror, fullspeed, instrcr, wang2021riff, wang2022odin, li2023accelerating}, and we integrate this metric in our heuristics.

2) RTL heuristics. This category contains metrics related to RTL hardware, including the number of {load} or {store} instructions ($\mathtt{ld\_st}$), floating point instructions ($\mathtt{fp}$), and jump instructions ($\mathtt{jp}$). According to our study of recent bugs in real-world cores \cite{boom, rocket}, e.g., Boom and Rocket, we find out that these instructions are the cause of multiple bugs. Therefore, we use these metrics in scoring the input instructions.

In conclusion, our heuristics calculation can be summarized as

\begin{equation}
	\label{equ_heu}
	heuristics = \frac{\omega *cov*speed + ld\_st*fp*jp}{len}
\end{equation}

In this equation, $\mathtt{\omega}$ is the proportion of basic metrics in the heuristics, which controls the weight between basic and RTL metrics. Based on this heuristics, we select input instructions with the highest score in the seed pool.

\subsection{Mutation}

In the design of DiFuzzRTL, the mutator can only add instructions to the inputs in mutation \cite{difuzzrtl}. In other words, the mutation operation will keep the length of input instructions increasing. However, in common fuzzing tools such as AFL, it is widely acknowledged that fuzzers should be equipped with various effective mutation strategies \cite{afl, mopt, fairfuzz}, e.g., ``dictionary'' (replacing part of the input instructions with tokens\footnote{Token means an item in the dictionary.}) and ``splice'' (splicing two inputs to generate one) in AFL.

We use several mutation strategies in \textsc{Instiller} to improve fuzzing. 1) The mutation strategy of DiFuzzRTL. The paper indicates its strategy is effective in fuzzing \cite{difuzzrtl}. Therefore, we retain it in \textsc{Instiller}. 2) Dictionary. In input instruction distillation, we get the distilled inputs, in which we can extract dictionary tokens to guide mutation. These tokens are effective in fuzzing and short in length. By replacing part of the original input instruction with these tokens, this ``dictionary'' mutation strategy is completed. 3) Insertion. We randomly insert new instructions to the inputs. 4) Deletion. Part of the input instruction is deleted, and the length of the input will decrease.

These four types of mutation strategies are similar to the mutation in binary fuzzing, but they are different. The input instruction of binary fuzzing can be encoded to sequences of ``1''s and ``0''s. The mutation on the sequence can be treated as mutating a simple string to enumerate all cases. Therefore, ``bitflip'' (flipping ``1'' to ``0'', or ``0'' to ``1'') in AFL is highly effective in discovering new paths \cite{mopt}. However, if there are illegal instructions, randomly mutating the input instructions of RTL can be often meaningless. Therefore, we use several techniques in the design of mutation strategies. For example, when inserting instructions, we use previously-used ones with higher probability and newly-generated ones with lower probability. Another example is in deletion, when deleting one instruction, the related instruction will also be deleted, e.g., a jump instruction and its target.

Furthermore, these four types of mutation in \textsc{Instiller} are invoked in different situations. They are chosen with different probability, which is calculated by weighted metrics. Every strategy is chosen with a different probability. If the condition is satisfied, the respective mutation strategy is selected. Otherwise, \textsc{Instiller} uses the original mutation. This can be concluded as

\begin{equation}
	\label{equ_mut}
	mutation = \left\{
	\begin{array}{ll}
		\mathtt{dictionary} & coverage\; decreases \\
		\mathtt{insertion} & len < l \\
		\mathtt{deletion} & len > l \\
		\mathtt{basic} & otherwise\\
	\end{array}
	\right.
\end{equation}

\section{Implementation}

We implement \textsc{Instiller} based on DiFuzzRTL. In general, the RTL instrumentation is finished with FIRRTL . We use Spike for ISA simulation, and cocotb for RTL simulation. The main fuzzing loop is implemented in Python.

In total, the implementation of \textsc{Instiller} can be divided into three parts. 1) The input instruction distillation process, which includes relationship extraction and VACO. We add this part in $\mathtt{Fuzzer.py}$, and there are about 500 lines of code in total. Additionally, we modify the main fuzzing loop to interact with VACO, including the starting and terminating conditions. 2) Interruptions and exceptions. We integrate multiple interruptions and exceptions into \textsc{Instiller}, as well as their priorities. $\mathtt{signature\_checker.py}$ is modified, where we focus on the special registers for the interruption and exception handling, e.g., $\mathtt{scause}$ and $\mathtt{mcause}$. This part contains about 550 lines of code. 3) Seed selection and mutation strategies. We implement the heuristics calculation and four mutation strategies in $\mathtt{mutator.py}$, which include about 700 lines of code. Both seed selection and mutation are integrated into the original process, which will cause no additional steps.

\section{Evaluation}

In our evaluation, we answer the following research questions:
\begin{itemize}
	\item \textbf{RQ1. }Can \textsc{Instiller} increase code coverage and shorten input instruction length?
	\item \textbf{RQ2. }Does \textsc{Instiller} have better vulnerability discovery ability than state-of-the-art RTL fuzzers?
	\item \textbf{RQ3. }What is the performance increase in execution speed of \textsc{Instiller}?
	\item \textbf{RQ4. }How does the VACO algorithm perform?
	\item \textbf{RQ5. }How do the techniques on interruptions and exceptions perform?
	\item \textbf{RQ6. }How do the seed selection and mutation strategies perform?
\end{itemize}

\subsection{Setup}

\begin{table}[htbp]
	\setlength{\abovecaptionskip}{-0.1cm}
	\caption{Information of the target CPUs, including ISA, number of pipelines, instruction width, and release time}
	\label{table_target}
	\centering
	\resizebox{0.8\columnwidth}{!}{
		\begin{tabular}{ccccc}
			\bottomrule
			\bfseries Targets & \bfseries ISA & \textbf{Pipeline} & \textbf{Width} &\textbf{Year}\\
			\toprule
			mor1kx & OpenRISC & 6-stage & 32-bit & 2013\\
			or1200 & OpenRISC & 5-stage & 32-bit & 2000\\
			Boom & RISC-V & 4-stage & 32-bit & 2017\\
			Rocket & RISC-V & 5-stage & 32-bit & 2016\\
			\bottomrule
		\end{tabular}
	}
\end{table}

\textbf{Targets. }In our experiments, the target RTL designs include mor1kx \cite{mor1kx}, or1200 \cite{or1200}, Boom \cite{boom}, and Rocket \cite{rocket}. These are popular RTL cores, and state-of-the-art papers used them in the experiments \cite{difuzzrtl, kande2022thehuzz}. Therefore, including these RTL designs in our evaluation demonstrates persuasiveness and representativeness. The detailed information is listed in Table \ref{table_target}.

\textbf{Compared tools. }To demonstrate the performance of \textsc{Instiller}, we compare it with DiFuzzRTL. DiFuzzRTL is one of the state-of-the-art CPU testing tools. We will compare \textsc{Instiller} with DiFuzzRTL in different aspects such as coverage\footnote{The source code of TheHuzz \cite{kande2022thehuzz} is not available. Therefore, it is not evaluated in this paper.}.

\textbf{Metrics. }We use coverage, the length of input instructions, the number of mismatches, and execution speed as the metrics in our evaluation. For every target CPU, we repeat the 24-hour fuzzing 10 times. In addition, we calculate the p values and ${\hat{A}_{12}}$ values of all the experiment results to eliminate the effect of randomness in fuzzing \cite{klees2018evaluating}. If the $p<0.05$ and ${\hat{A}_{12}} > 0.5$, then this specific comparison shows a statistically significant difference.


\subsection{Evaluation on Coverage} \label{sec_eva_cov}

\begin{table}[htbp]
	\setlength{\abovecaptionskip}{-0.1cm}
	\caption{Evaluation on coverage of \textsc{Instiller} and DiFuzzRTL, where the value in the bracket denotes the increase or decrease compared with the competitors}
	\label{table_cov_sim}
	\centering
	\resizebox{\columnwidth}{!}{
		\begin{tabular}{ccccc}
			\bottomrule
			\bfseries Targets & \bfseries \textsc{Instiller} & \textbf{DiFuzzRTL} & \textbf{p value} & \textbf{${\hat{A}_{12}}$}\\
			\toprule
			mor1kx & 201288.7(+32.6\%) & 151840.5 & 5.11$*10^{-4}$ & 1.0\\
			or1200 & 269504.8(\textbf{+36.5\%}) & 197503.9 & 6.84$*10^{-5}$ & 1.0\\
			Boom & 547433.5(+28.8\%) & 425183.7 & 9.13$*10^{-5}$ & 1.0 \\
			Rocket & 101489.6(+11.9\%) & 90715.3 & 8.98$*10^{-5}$ & 1.0\\
			\toprule
			Average & 279929.15(+29.4\%) & 216310.85 & 1.90$*10^{-4}$ &1.0 \\
			\bottomrule
		\end{tabular}
	}
\end{table}

Table \ref{table_cov_sim} is the coverage results of \textsc{Instiller} and DiFuzzRTL. According to the table, \textsc{Instiller} outperforms DiFuzzRTL in all the targets, and all the coverage increase is more than 11\%. In all the comparisons with DiFuzzRTL, the p values are less than 0.05, and the ${\hat{A}_{12}}$ values are greater than 0.5. The results indicate statistically significant differences. In the comparison in or1200, \textsc{Instiller} has 36.5\% more coverage than DiFuzzRTL, which is the greatest difference among all the results. On average, \textsc{Instiller} reaches 29.4\% more coverage, and this result also has a statistically significant difference.

\begin{figure}[htbp]
	\setlength{\abovecaptionskip}{-0.2cm}
	\centering
	\includegraphics[width=0.9\columnwidth]{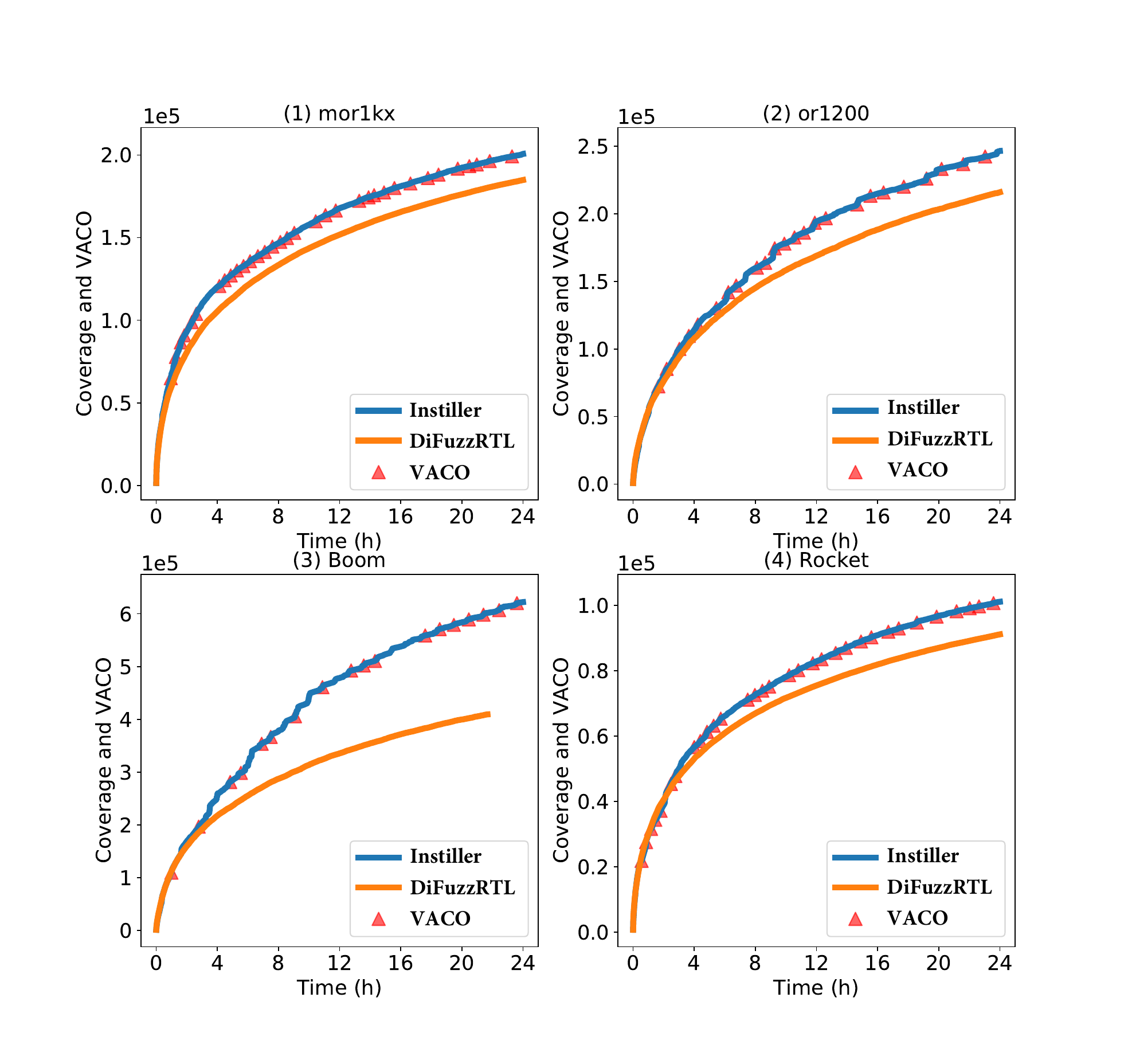}
	\caption{{Coverage and the effectiveness of VACO over time, where the X-axis is the time, and the Y-axis is the coverage.}}
	\label{fig_cov_VACO}
\end{figure}

Figure \ref{fig_cov_VACO} shows the coverage growing over time of \textsc{Instiller} and DiFuzzRTL. We record the results within 24 hours of fuzzing. According to the figure, in all the targets, the coverage of \textsc{Instiller} is greater, and the growth of coverage in \textsc{Instiller} is faster than DiFuzzRTL.

The results in Table \ref{table_cov_sim} and Figure \ref{fig_cov_VACO} demonstrate the effectiveness of input instruction distillation. The internal reason is that we distill inputs based on the coverage performance of input instructions, e.g., Line 1 in Algorithm \ref{algo_relation} sorts the inputs based on coverage.

In conclusion, according to these experiment results, \textsc{Instiller} has better coverage exploration ability than DiFuzzRTL.

\subsection{Evaluation on Input Instruction Length} \label{sec_length}

\begin{table}[htbp]
	\setlength{\abovecaptionskip}{-0.1cm}
	\caption{Evaluation on input instruction length of \textsc{Instiller} and DiFuzzRTL, where the value in the bracket denotes the increase or decrease compared with the competitors}
	\label{table_len_sim}
	\centering
	\resizebox{\columnwidth}{!}{
		\begin{tabular}{ccccc}
			\bottomrule
			\bfseries Targets & \bfseries \textsc{Instiller} & \textbf{DiFuzzRTL} & \textbf{p value} & \textbf{${\hat{A}_{12}}$}\\
			\toprule
			mor1kx & 451.39(-77.6\%) & 2018.21 & 9.56$*10^{-5}$ & 1.0\\
			or1200 & 385.63(-80.6\%) & 1986.70 & 7.55$*10^{-5}$ & 1.0\\
			Boom & 486.75(-74.1\%) & 1882.80 & 9.13$*10^{-5}$ & 1.0\\
			Rocket & 420.67(\textbf{-83.6\%}) & 2557.67 & 8.98$*10^{-5}$ & 1.0\\
			\toprule
			Average & 436.11(-79.3\%) & 2111.34 & 8.80$*10^{-5}$ & 1.0\\
			\bottomrule
		\end{tabular}
	}
\end{table}

Table \ref{table_len_sim} shows the input instruction length during fuzzing of \textsc{Instiller} and DiFuzzRTL, respectively. In all the target CPU cores, \textsc{Instiller} shortens the length of input instructions, and all the decrease is more than 74\%. Especially in the Rocket core, the decrease of length reaches the maximum of 83.6\%. In addition, all the comparisons with DiFuzzRTL have statistically significant differences. The average input instruction length of \textsc{Instiller} is 79.3\% shorter than DiFuzzRTL. These results indicate the effectiveness of VACO, which significantly shortens the input instruction length.

\vspace{-0.0cm}
\begin{figure}[htbp]
	\setlength{\abovecaptionskip}{-0.5cm}
	\centering
	\includegraphics[width=\columnwidth]{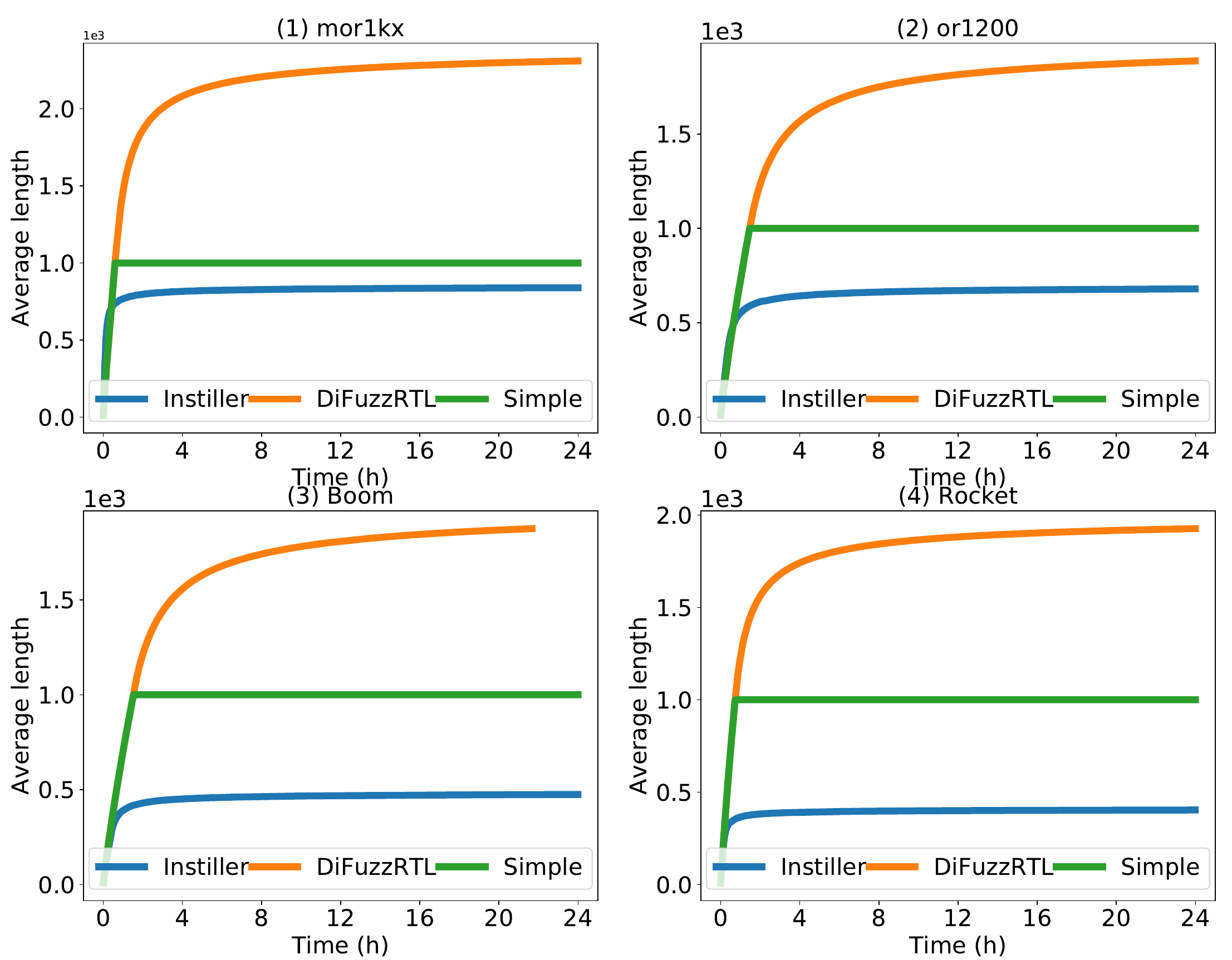}
	\caption{{Input instruction length over time, where the X-axis is the time, and the Y-axis is the average length.}}
	\label{fig_len_ave}
\end{figure}
\vspace{-0.0cm}

Figure \ref{fig_len_ave} is the result of input instruction length over time. In the 24 hours, compared with DiFuzzRTL, \textsc{Instiller} shortens the length.

The results in this section show a significant difference between input instruction length in \textsc{Instiller} and DiFuzzRTL. Decreasing the input length is the main focus of this work, and the mechanism in VACO contributes to the results. Moreover, the results of DiFuzzRTL in Table \ref{table_cov_sim} and Figure \ref{fig_len_ave} demonstrate the statement in Section \ref{sec_intro} that coverage does not increase proportionally to input length.

$\bullet$ Therefore, based on the results of coverage and input instruction length, we can answer RQ1: \textit{\textsc{Instiller} can increase code coverage and shorten input instruction length at the same time.}

\subsection{Evaluation on Vulnerability Detection}

\begin{table}[htbp]
	\setlength{\abovecaptionskip}{-0.1cm}
	\caption{Number of mismatches of \textsc{Instiller} and DiFuzzRTL, where the value in the bracket denotes the increase or decrease compared with the competitors}
	\label{table_bug}
	\centering
	\resizebox{0.9\columnwidth}{!}{
		\begin{tabular}{ccccc}
			\bottomrule
			\bfseries Targets & \bfseries \textsc{Instiller} & \textbf{DiFuzzRTL} & \textbf{p value} & \textbf{${\hat{A}_{12}}$}\\
			\toprule
			mor1kx & 110.1({-3.9\%}) & 120.9 & 0.98 & 0.2\\
			or1200 & 598.3(+6.7\%) & 560.9 & 0.001 & 1.0\\
			Boom & 5546.0(\textbf{+18.8\%}) & 4666.6 & 0.01 & 1.0\\
			Rocket & 33.3({+14.0\%}) & 29.2 & 0.06 & 0.7\\
			\toprule
			Average & 1573.9(+17.0\%) & 1344.4 & 0.26 & 0.73\\
			\bottomrule
		\end{tabular}
	}
\end{table}

Table \ref{table_bug} shows the number of mismatches of differential testing. In the design of \textsc{Instiller} and DiFuzzRTL, if the output of ISA simulation is different from RTL execution, i.e., a mismatch, a potential bug is detected. It is rational to use the number of mismatches to demonstrate the vulnerability detection ability of the fuzzing tools. In the table, except for mor1kx, \textsc{Instiller} outperforms DiFuzzRTL in all the targets. The greatest improvement is in Boom core, with an increase of 18.8\%. On average, \textsc{Instiller} also outperforms DiFuzzRTL in vulnerability discovery, and it detects 16.9\% more mismatches.

Furthermore, we manually investigate why DiFuzzRTL outperforms \textsc{Instiller} in mor1kx. In Table \ref{table_cov_sim}, \textsc{Instiller} has more coverage. Our investigation shows the fuzzers cover different parts of mor1kx. Moreover, by looking into the code of mor1kx, we find out that the hardware circuits covered by \textsc{Instiller} cannot trigger as many mismatches as DiFuzzRTL. These factors lead to the results in Table \ref{table_bug}.

$\bullet$ Therefore, we can answer RQ2: \textit{\textsc{Instiller} has better vulnerability detection ability than DiFuzzRTL.}

\subsection{Evaluation on Execution Speed}

\begin{table}[htbp]
	\setlength{\abovecaptionskip}{-0.1cm}
	\caption{Execution speed per second of \textsc{Instiller} and DiFuzzRTL, where the values in the bracket denote the increase or decrease compared with the competitors}
	\label{table_speed_sim}
	\centering
	\resizebox{0.8\columnwidth}{!}{
		\begin{tabular}{ccccc}
			\bottomrule
			\bfseries Targets & \bfseries \textsc{Instiller} & \textbf{DiFuzzRTL} & \textbf{p value} & \textbf{${\hat{A}_{12}}$}\\
			\toprule
			mor1kx & 0.27({+8.0\%}) & 0.25 & 0.99 & 0.15\\
			or1200 & 0.40(+5.3\%) & 0.38 & 0.98 & 0.21\\
			Boom & 0.25(+4.1\%) & 0.24 & 0.99 & 0.10\\
			Rocket & 0.36(\textbf{+9.1\%}) & 0.33 & 0.99 & 0.20\\
			\toprule
			Average & 0.32(+6.7\%) & 0.30 & 0.99 & 0.17\\
			\bottomrule
		\end{tabular}
	}
\end{table}

Table \ref{table_speed_sim} shows the execution speed of \textsc{Instiller} and DiFuzzRTL. Execution speed is the result of dividing the number of executions by the time. In general, \textsc{Instiller} runs faster than DiFuzzRTL. The greatest difference is in Rocket, which is 9.1\%. On average, the performance increase in \textsc{Instiller} is 6.7\%, which shows the effectiveness of distilled input instructions. Shorter inputs require fewer CPU cycles to execute.

In Section \ref{sec_length}, the input instruction length of \textsc{Instiller} is 79.3\% shorter on average. Intuitively, the performance increase should have been more than 6.7\%. We investigate the cause of this result. First, there is no direct correspondence between the input length and execution speed of fuzzing. Having 79.3\% short length does not mean executing 79.3\% faster. Second, there is a performance overhead in the process of relationship extraction and VACO. The relationship extraction enumerates all the executed input instructions, and VACO is also an iterative process. Therefore, the performance overhead is unavoidable to finish these processes.

\begin{table}[htbp]
	\setlength{\abovecaptionskip}{-0.1cm}
	\caption{Performance overhead of relationship extraction and VACO, where the values in the bracket denote the increase or decrease compared with the competitors}
	\label{table_re_vaco_oh}
	\centering
	\begin{threeparttable}
		\resizebox{\columnwidth}{!}{
			\begin{tabular}{lcccr}
				\bottomrule
				\bfseries  & \bfseries \textsc{Instiller} & \textbf{\textsc{Instiller}$\scriptscriptstyle \rm^{-R}$} & \textbf{\textsc{Instiller}$\scriptscriptstyle \rm^{-V}$} & \textbf{\textsc{Instiller}$\scriptscriptstyle \rm^{-RV}$}\\
				\toprule
				Speed & 0.32 & 0.34(+6.3\%) & 0.29(-9.4\%) & 0.30(-6.3\%)\\
				\bottomrule
			\end{tabular}
		}
		\begin{tablenotes}
			\item[1] \textsc{Instiller}$\scriptscriptstyle \rm^{-R}$ denotes \textsc{Instiller} without relationship extraction.
			\item[2] \textsc{Instiller}$\scriptscriptstyle \rm^{-V}$ denotes \textsc{Instiller} without VACO.
			\item[3] \textsc{Instiller}$\scriptscriptstyle \rm^{-RV}$ denotes \textsc{Instiller} without relationship extraction\\ and VACO.
		\end{tablenotes}
	\end{threeparttable}
\end{table}

In addition, we conduct extra experiments to show their specific overhead. Table \ref{table_re_vaco_oh} shows the respective average speed of four different configurations of \textsc{Instiller}. When relationship extraction is disabled, the fuzzer can run 6.3\% faster than the original \textsc{Instiller}. However, investigating coverage data shows that this configuration has less coverage than \textsc{Instiller}. Therefore, disabling relationship extraction has an effect on speed and coverage. If there is no VACO in \textsc{Instiller}, the execution speed decreases by 9.4\%. The reason is that disabling VACO means input distillation is disabled, which cannot utilize shorter input instructions. Besides, the execution of relationship extraction in \textsc{Instiller}$^{\scriptscriptstyle \rm -V}$ has overhead. The average speed of \textsc{Instiller}$^{\scriptscriptstyle \rm -RV}$ is the same as DiFuzzRTL in Table \ref{table_speed_sim}, which means other approaches of \textsc{Instiller}, e.g., seed selection and mutation, cause negligible performance overhead.

$\bullet$ We can answer RQ3: \textit{The performance increase of \textsc{Instiller} is 6.7\% on average.}

%
%
%

\subsection{Results of VACO}

\begin{figure}[htbp]
	\setlength{\abovecaptionskip}{-0.5cm}
	\centering
	\includegraphics[width=\columnwidth]{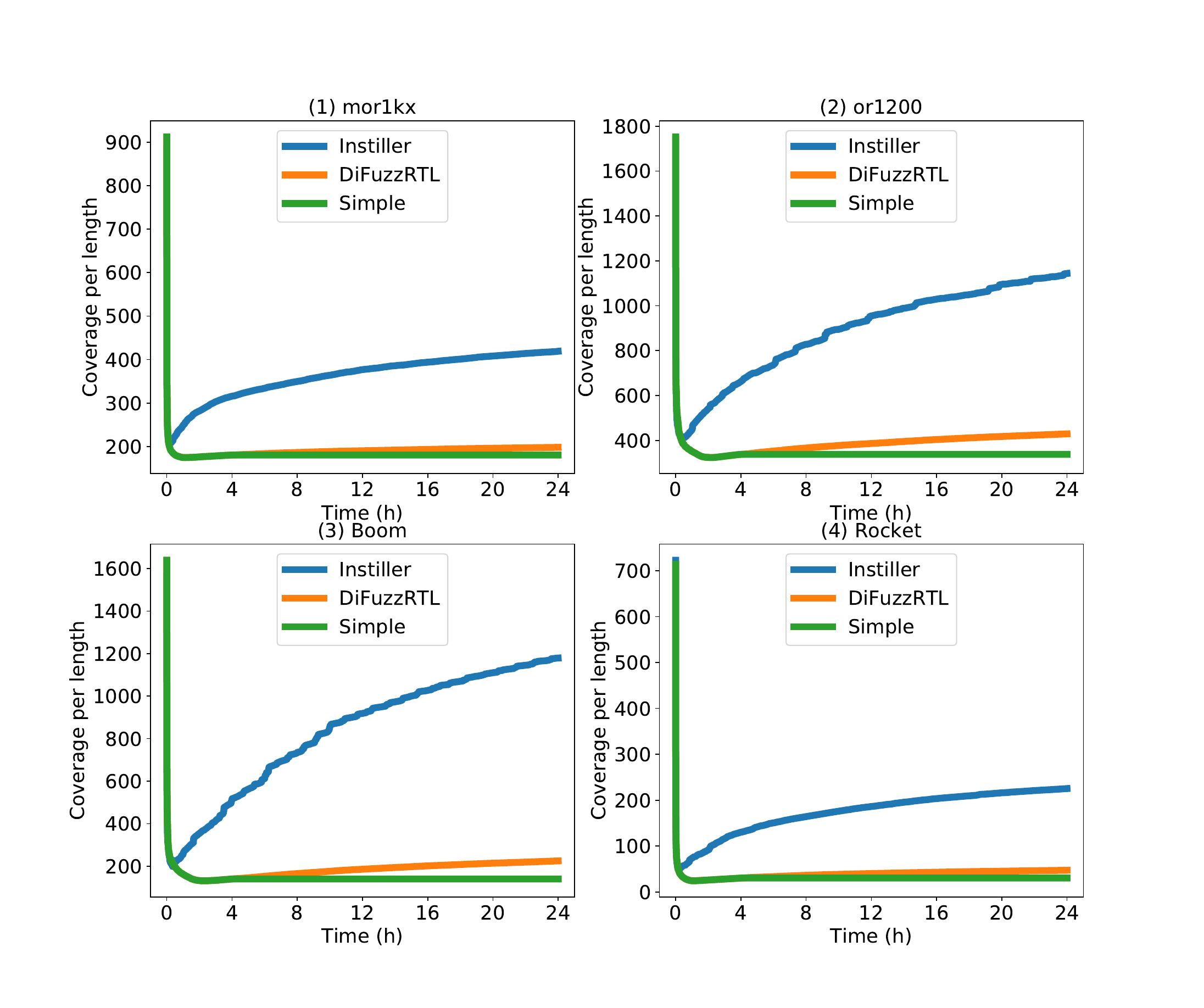}
	\caption{{Coverage divided by input instruction length over time, where the X-axis denotes time, and the Y-axis denotes coverage divided by length.}}
	\label{fig_cov_per_len}
\end{figure}
\vspace{-0.0cm}

Figure \ref{fig_cov_per_len} shows the results of coverage divided by input instruction length over time of \textsc{Instiller} and DiFuzzRTL. Both of the lines experience a high start in the figure. The reason is that the first input instruction length is always ``1'', and executing the first input brings about a 700 to 1800 coverage increase. Regardless of the high start, \textsc{Instiller} has higher coverage per length than DiFuzzRTL. This result indicates the input instructions in \textsc{Instiller} are more effective in discovering new coverage. The mechanisms of relationship extraction and VACO ensure this result.

Figure \ref{fig_cov_VACO} shows the coverage results of \textsc{Instiller} and DiFuzzRTL, and how VACO affects coverage in the 24 hours. The red circles in the figure indicate the enabling of VACO. Every time VACO is enabled, the coverage increases more rapidly than DiFuzzRTL.

\begin{table}[htbp]
	\setlength{\abovecaptionskip}{-0.1cm}
	\caption{Results of coverage and input instruction length of different configurations of \textsc{Instiller}, where the values in the bracket denote the increase or decrease compared with the competitors}
	\label{table_cl_rv}
	\centering
	\begin{threeparttable}
		\resizebox{\columnwidth}{!}{
			\begin{tabular}{lcccr}
				\bottomrule
				\bfseries  & \bfseries \textsc{Instiller} & \textbf{\textsc{Instiller}$\scriptscriptstyle \rm^{-R}$} & \textbf{\textsc{Instiller}$\scriptscriptstyle \rm^{-V}$} & \textbf{\textsc{Instiller}$\scriptscriptstyle \rm^{-RV}$} \\
				\toprule
				Coverage & 279929.2 & 248663.5(-11.2\%) & 255439.2(-8.7\%) & 217197.5(-22.4\%)\\
				Length & 436.1 & 498.4(+14.3\%) & 2185.9(+401.2\%) & 2254.8(+417.0\%)\\
				\bottomrule
			\end{tabular}
		}
		\begin{tablenotes}
			\item[1] \textsc{Instiller}$\scriptscriptstyle \rm^{-R}$ denotes \textsc{Instiller} without relationship extraction.
			\item[2] \textsc{Instiller}$\scriptscriptstyle \rm^{-V}$ denotes \textsc{Instiller} without VACO.
			\item[3] \textsc{Instiller}$\scriptscriptstyle \rm^{-RV}$ denotes \textsc{Instiller} without relationship extraction\\ and VACO.
		\end{tablenotes}
	\end{threeparttable}
\end{table}

Table \ref{table_cl_rv} shows the results of coverage and input instruction length of four different configurations of \textsc{Instiller}. For coverage, disabling relationship extraction and VACO will have a negative effect. With a 22.4\% decrease, \textsc{Instiller}$^{\scriptscriptstyle \rm -RV}$ has the least coverage, which is almost the same as the result of DiFuzzRTL in Table \ref{table_cov_sim}. For input instruction length, the VACO algorithm has the most effect. Disabling VACO alone causes a 417.0\% length increase, which is also close to the result of DiFuzzRTL in Table \ref{table_len_sim}.

Besides, we compare VACO with a simpler method which sets an upper limit for the length and discards seeds exceeding this limit \footnote{``Simple" strategy is the version of \textsc{INSTILLER} that replaces VACO with an upper-bound-limiting strategy.}. The results are shown in Figure \ref{fig_len_ave} and Figure \ref{fig_cov_per_len}. The upper limit is set to 1,000 in our evaluation. Figure \ref{fig_len_ave} denotes the average length over time of this simple strategy. In this figure, the average length of {Simple} grows as the fuzzing process continues. When the length reaches the limit, it stays unchanged until the end of fuzzing.

Moreover, as shown in Figure \ref{fig_cov_per_len}, the result of {Simple} experiences a high start in the beginning, which is similar to \textsc{Instiller} and DiFuzzRTL. However, as the fuzzing process continues, the coverage per length of {Simple} stays at a low value compared with \textsc{Instiller}, and the value remains unchanged several hours after the start of fuzzing. Though simply setting an upper limit for the length and discarding seeds exceeding this limit can control the length of the input sequence, there is no benefit in improving coverage. Our VACO algorithm surpasses the basic upper-bound-limiting strategy by controlling the input length and increasing coverage at the same time.

Furthermore, we conduct experiments comparing the bug-finding performance between \textsc{Instiller} and \textsc{Instiller} without VACO (denoted as \textsc{Instiller$\scriptscriptstyle \rm^{-V}$}). As shown in Table \ref{table_bug_VACO}, \textsc{Instiller} outperforms \textsc{Instiller$\scriptscriptstyle \rm^{-V}$} in all the target CPUs. On average, \textsc{Instiller} has 12.5\% more mismatches, demonstrating better vulnerability discovery ability of \textsc{Instiller}. The reason behind these results is mainly due to the coverage of the fuzzing process. As shown in Table \ref{table_cl_rv}, \textsc{Instiller$\scriptscriptstyle \rm^{-V}$} has 8.7\% less coverage than \textsc{Instiller}. Covering more code ensures \textsc{Instiller} to find more bugs.

\begin{table}[htbp]
	\setlength{\abovecaptionskip}{-0.1cm}
	\caption{Number of mismatches of \textsc{Instiller} and \textsc{Instiller} without VACO (denoted as \textsc{Instiller$\scriptscriptstyle \rm^{-V}$}), where the values in the bracket denote the increase or decrease compared with the competitors}
	\label{table_bug_VACO}
	\centering
	\resizebox{0.9\columnwidth}{!}{
		\begin{tabular}{ccccc}
			\bottomrule
			\bfseries Targets & \bfseries \textsc{Instiller} & \textbf{\textsc{Instiller$\scriptscriptstyle \rm^{-V}$}} & \textbf{p value} & \textbf{${\hat{A}_{12}}$}\\
			\toprule
			mor1kx & 110.1({+7.4\%}) & 102.5 & 0.04 & 0.7\\
			or1200 & 598.3(+6.5\%) & 561.7 & 0.002 & 1.0\\
			Boom & 5546.0(\textbf{+13.3\%}) & 4897.1 & 0.01 & 1.0\\
			Rocket & 33.3({+10.3\%}) & 30.2 & 0.045 & 1.0\\
			\toprule
			Average & 1573.9(+12.5\%) & 1397.9 & 0.022 & 0.93\\
			\bottomrule
		\end{tabular}
	}
\end{table}

$\bullet$ Therefore, we can answer RQ4: \textit{Experiment results show the effectiveness of VACO, together with relationship extraction, which can increase code coverage, shorten input instruction length, and find more bugs at the same time.}

\subsection{Results of Multiple Interruptions and Exceptions}

\vspace{-0.0cm}
\begin{figure}[htbp]
	\setlength{\abovecaptionskip}{-0.2cm}
	\centering
	\includegraphics[width=0.9\columnwidth]{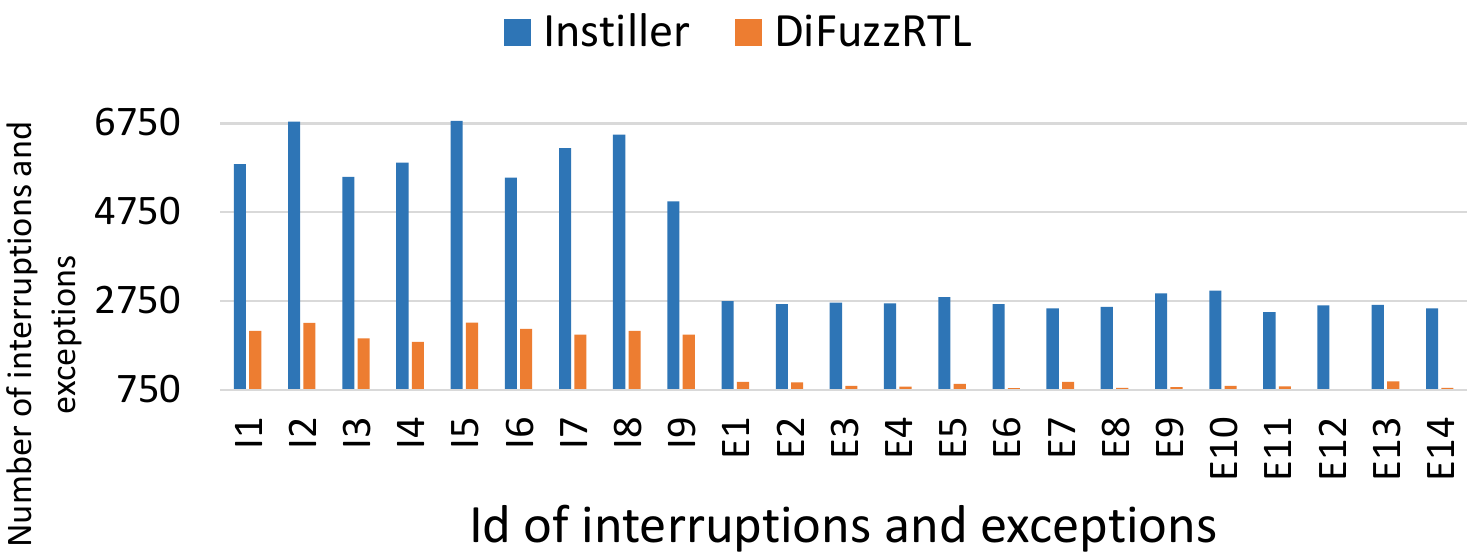}
	\caption{{Number of interruptions and exceptions of \textsc{Instiller} and DiFuzzRTL, where the X-axis denotes ID, and the Y-axis denotes the number.}}
	\label{fig_intr_num}
\end{figure}
\vspace{-0.0cm}

We collect 9 interruptions (\textit{I1} - \textit{I9}) and 14 exceptions (\textit{E1} - \textit{E14}) from \cite{waterman2014risc}. In this part of the evaluation, we insert interruptions and exceptions into the input instructions with 50\% probability. Moreover, we set the limit of interrupts and exceptions to three, respectively, to show the effectiveness of multiple ones. However, this does not mean \textsc{Instiller} can only support three interruptions or exceptions. The limit can be configured as needed.

Figure \ref{fig_intr_num} shows the number of interruptions and exceptions of \textsc{Instiller} and DiFuzzRTL. The number of interruptions and exceptions of \textsc{Instiller} are more than DiFuzzRTL. Multiple ones are inserted into the input instructions, and the chance of triggering new states and bugs is higher in \textsc{Instiller}.

In addition, we compare \textsc{Instiller} with DiFuzzRTL by considering the priorities of interruptions and exceptions. In the design of RISC-V, the interruptions and exceptions have fixed priorities. For example, ``I2'' is higher than ``I0'', and ``E3'' is higher than ``E0''. Therefore, different combinations of them indicate different combinations of priorities. In this part, we define \textit{interruption state transition (IST)} and \textit{exception state transition (EST)} to describe different combinations of interruptions and exceptions. For example, there are ``I3'' and ``I7'',  and ``E2'' in an input instruction. The IST of this input instruction is 1 ((3 $\ll$ 1) XOR 7), and EST is 4 ((2 $\ll$ 1) XOR 0). Here, we use $(ID_1$ $\ll$ 1) XOR $ID_2)$ as a hash method to represent IST or EST. In our definition, the more different values of IST or EST, the more different states a fuzzer can reach.

\begin{figure}[htbp]
	\setlength{\abovecaptionskip}{-0.2cm}
	\centering
	\includegraphics[width=0.8\columnwidth]{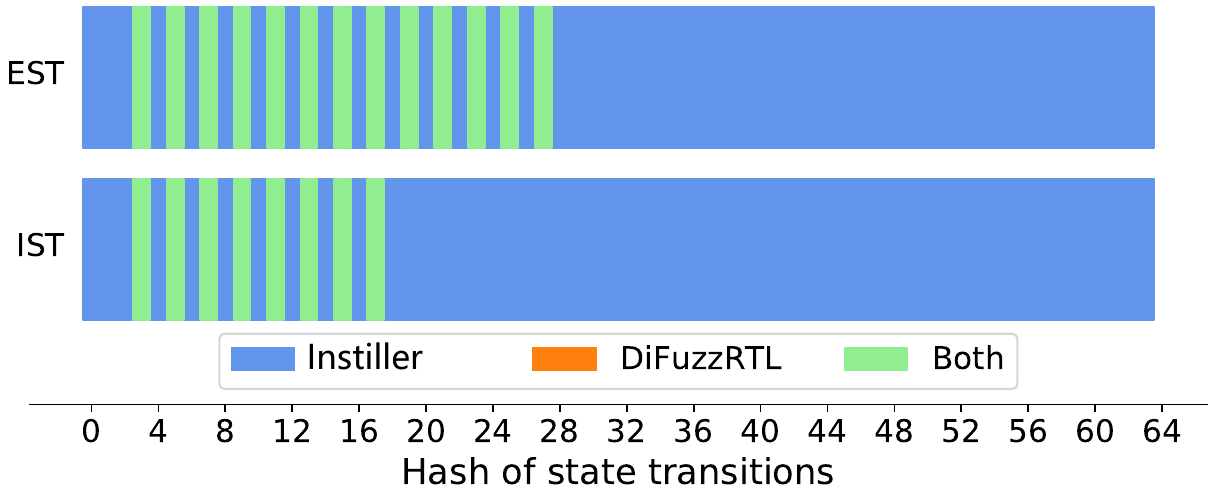}
	\caption{{Distribution of IST and EST of \textsc{Instiller} and DiFuzzRTL.}}
	\label{fig_ist_est}
\end{figure}

\vspace{-0.4cm}

Figure \ref{fig_ist_est} shows the distribution of state transitions of interruptions and exceptions. The IST of \textsc{Instiller} ranges from 0 to 64, while that of DiFuzzRTL ranges from 2 to 18. The EST result is similar. The states \textsc{Instiller} can reach are more than DiFuzzRTL. This result indicates that by considering multiple interruptions and exceptions with their combinations, \textsc{Instiller} can trigger more states in fuzzing, and therefore, it has a higher chance to discover bugs. Note that in the figure, the result of ``DiFuzzRTL'' is covered by ``Both'', so we cannot see the result of it.

\begin{table}[htbp]
	\setlength{\abovecaptionskip}{-0.1cm}
	\caption{Number of mismatches of \textsc{Instiller} and \textsc{Instiller} without multiple interruptions and exceptions (denoted as \textsc{Instiller$\scriptscriptstyle \rm^{-IE}$}), where the values in the bracket denote the increase or decrease compared with the competitors}
	\label{table_bug_IE}
	\centering
	\resizebox{0.8\columnwidth}{!}{
		\begin{tabular}{ccccc}
			\bottomrule
			\bfseries Targets & \bfseries \textsc{Instiller} & \textbf{\textsc{Instiller$\scriptscriptstyle \rm^{-IE}$}} & \textbf{p value} & \textbf{${\hat{A}_{12}}$}\\
			\toprule
			mor1kx & 110.1({+2.8\%}) & 107.1 & 0.015 & 0.85\\
			or1200 & 598.3(+3.7\%) & 577.2 & 0.01 & 0.9\\
			Boom & 5546.0({+2.5\%}) & 5411.8 & 0.03 & 1.0\\
			Rocket & 33.3(\textbf{+7.1\%}) & 31.1 & 0.02 & 1.0\\
			\toprule
			Average & 1573.9(+2.6\%) & 1531.8 & 0.019 & 0.94\\
			\bottomrule
		\end{tabular}
	}
\end{table}

\vspace{-0.4cm}

Moreover, Table \ref{table_bug_IE} compares the bug-finding performance between \textsc{Instiller} and \textsc{Instiller} without multiple interruptions and exceptions (denoted as \textsc{Instiller$\scriptscriptstyle \rm^{-IE}$}). As the table illustrates, \textsc{Instiller} surpasses \textsc{Instiller$\scriptscriptstyle \rm^{-IE}$} in all the CPUs, reaching up to 7.1\% in Rocket. On average, \textsc{Instiller} also outperforms the competitor by 2.6\%. Besides, all the p values are less than 0.05, and all the ${\hat{A}_{12}}$ values are greater than 0.5, indicating all the results have significant differences.

We investigate the source code and find out that some mismatches in Table \ref{table_bug_IE} reside in the PLIC, nested interruption handling, and exception handling of CPU implementations, as listed in Section \ref{sec_2_A} and Figure \ref{fig_multi_intr_cov}. For example, we witness a store page fault exception handling in Rocket core using \textsc{Instiller}, while we cannot reproduce it with other fuzzers or \textsc{Instiller$\scriptscriptstyle \rm^{-IE}$}. This result verifies our investigation in Section \ref{sec_2_A} that enabling multiple interruptions and exceptions is effective to CPU fuzzing.

$\bullet$ Therefore, we can answer RQ5: \textit{The techniques of multiple interruptions and exceptions are effective.}

\subsection{Effectiveness of Seed Selection and Mutation}

\begin{table}[htbp]
	\setlength{\abovecaptionskip}{-0.1cm}
	\caption{Coverage and length comparison of \textsc{Instiller} with other configurations, where the values in the bracket denote the increase or decrease compared with the competitors}
	\label{table_seed_mut}
	\centering
	\begin{threeparttable}
		\resizebox{\columnwidth}{!}{
			\begin{tabular}{lcccr}
				\bottomrule
				\bfseries Targets & \bfseries \textsc{Instiller} & \textbf{\textsc{Instiller}$\scriptscriptstyle \rm^{-S}$} & \textbf{\textsc{Instiller}$\scriptscriptstyle \rm^{-M}$} & \textbf{\textsc{Instiller}$\scriptscriptstyle \rm^{-SM}$}\\
				\toprule
				Coverage & 279929.15 & 267754.12(-4.3\%) & 259643.88(-7.2\%) & 255386.9(-8.8\%)\\
				Length & 436.11 & 467.08(+7.1\%) & 455.86(+4.5\%) & 478.1(+9.6\%) \\
				\bottomrule
			\end{tabular}
		}
		\begin{tablenotes}
			\item[1] \textsc{Instiller}$\scriptscriptstyle \rm^{-S}$ denotes \textsc{Instiller} without seed selection strategy.
			\item[2] \textsc{Instiller}$\scriptscriptstyle \rm^{-M}$ denotes \textsc{Instiller} without mutation strategy.
			\item[3] \textsc{Instiller}$\scriptscriptstyle \rm^{-SM}$ denotes \textsc{Instiller} without seed selection and\\ mutation strategies.
		\end{tablenotes}
	\end{threeparttable}
\end{table}

Table \ref{table_seed_mut} shows the results of coverage and length of different configurations of \textsc{Instiller}. By using Equation \ref{equ_heu}, the fuzzing process is guided towards increasing coverage. Therefore, \textsc{Instiller} has 4.3\% more coverage than \textsc{Instiller}$\scriptscriptstyle \rm^{-S}$. Similarly, Equation \ref{equ_mut} uses dictionary mutation to mitigate coverage decrease, and \textsc{Instiller}$\scriptscriptstyle \rm^{-M}$ has 7.2\% less coverage when the mutation strategy is disabled. The coverage of \textsc{Instiller}$\scriptscriptstyle \rm^{-SM}$ is more than DiFuzzRTL in Table \ref{table_cov_sim}. This result indicates the effectiveness of other techniques in \textsc{Instiller}, e.g., the VACO algorithm.

For input instruction length, the result of \textsc{Instiller}$\scriptscriptstyle \rm^{-S}$ is 7.1\% greater than \textsc{Instiller}. In Equation \ref{equ_heu}, we use heuristics divided by length, aiming to select the shortest input instruction. Therefore, the seed selection strategy also helps shorten input instruction length. In our mutation strategy, the length of the input instruction is controlled by insertion and deletion. The result of \textsc{Instiller}$\scriptscriptstyle \rm^{-M}$ is 4.5\% greater than \textsc{Instiller}, indicating the effectiveness of the mutation strategy. \textsc{Instiller}$\scriptscriptstyle \rm^{-SM}$ has a shorter length than DiFuzzRTL in Table \ref{table_len_sim}, proving the huge effect of distillation on input instruction length.

Moreover, we conduct experiments to compare \textsc{Instiller} with \textsc{Instiller} without seed selection and mutation (denoted as \textsc{Instiller$\scriptscriptstyle \rm^{-SM}$}), which is illustrated in Table \ref{table_bug_SM}. \textsc{Instiller} outperforms \textsc{Instiller$\scriptscriptstyle \rm^{-SM}$} in all the tested CPUs. Especially in Rocket, the performance difference reaches the maximum of 7.8\%. On average, \textsc{Instiller} surpasses \textsc{Instiller$\scriptscriptstyle \rm^{-SM}$} by 3.0\%, and all the differences are significant according to the p values and ${\hat{A}_{12}}$ values. Referring to Equation \ref{equ_heu} and Equation \ref{equ_mut}, we add $cov$ in the heuristics in seed selection and mutation. Covering more parts of the code contributes to the bug-finding performance of \textsc{Instiller}.

\begin{table}[htbp]
	\setlength{\abovecaptionskip}{-0.1cm}
	\caption{Number of mismatches of \textsc{Instiller} and \textsc{Instiller} without seed selection and mutation (denoted as \textsc{Instiller$\scriptscriptstyle \rm^{-SM}$}), where the values in the bracket denote the increase or decrease compared with the competitors}
	\label{table_bug_SM}
	\centering
	\resizebox{0.8\columnwidth}{!}{
		\begin{tabular}{ccccc}
			\bottomrule
			\bfseries Targets & \bfseries \textsc{Instiller} & \textbf{\textsc{Instiller$\scriptscriptstyle \rm^{-SM}$}} & \textbf{p value} & \textbf{${\hat{A}_{12}}$}\\
			\toprule
			mor1kx & 110.1({+5.6\%}) & 104.3 & 0.002 & 1.0\\
			or1200 & 598.3(+5.1\%) & 569.3 & 0.01 & 1.0\\
			Boom & 5546.0({+2.7\%}) & 5399.3 & 0.001 & 1.0\\
			Rocket & 33.3(\textbf{+7.8\%}) & 30.9 & 0.015 & 1.0\\
			\toprule
			Average & 1573.9(+3.0\%) & 1525.95 & 0.007 & 1.0\\
			\bottomrule
		\end{tabular}
	}
\end{table}

$\bullet$ Therefore, we can answer RQ6: \textit{The seed selection and mutation strategies are also effective in increasing coverage, shortening input length, and detecting bugs.}

\subsection{Comparison with Other Hardware Fuzzers}

\begin{figure}[htbp]
	\setlength{\abovecaptionskip}{-0.05cm}
	\centering
	\includegraphics[width=\columnwidth]{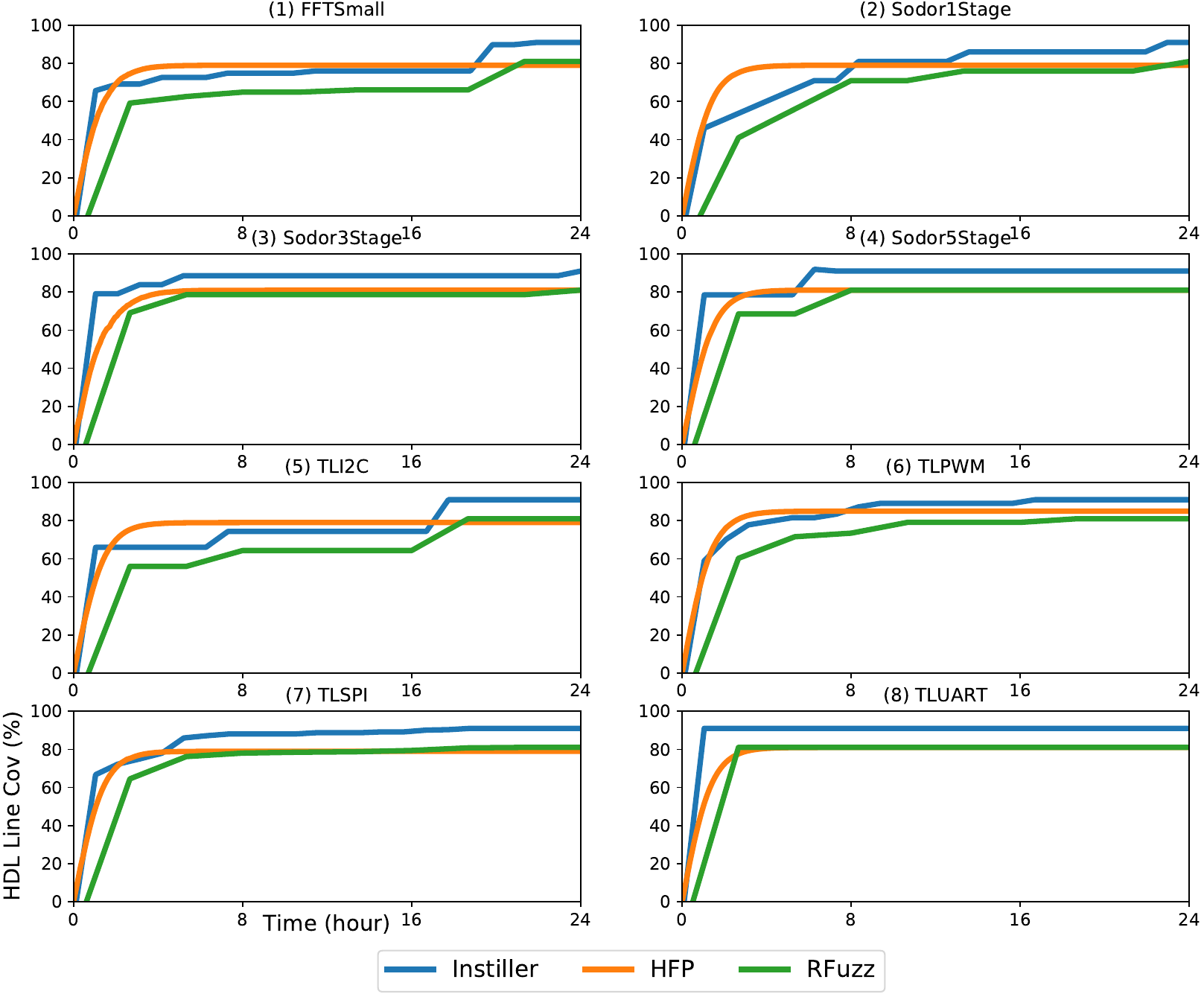}
	\caption{{HDL line coverage of \textsc{Instiller}, HFP, and RFuzz over time, where the X-axis denotes time, and the Y-axis denotes coverage.}}
	\label{fig_hdl}
\end{figure}

Despite using differential testing, there are other types of hardware fuzzers. In this part, we compare \textsc{Instiller} with Hardware Fuzzing Pipeline (HFP) \cite{trippel2022fuzzing} and RFuzz \cite{rfuzz}. Both HFP and RFuzz detect bugs without differential testing. There are eight experiment targets, including FFTSmall, Sodor1Stage, Sodor3Stage, Sodor5Stage, TLI2C, TLPWM, TLSPI, and TLUART. These targets are collected from the experiments of HFP and RFuzz. We use HDL line coverage as the coverage metric to show the results. As Figure \ref{fig_hdl} shows, \textsc{Instiller} has more HDL line coverage than HFP and RFuzz. The internal reason for this result is that we propose instruction distillation based on coverage in \textsc{Instiller}. The distilled input instructions are short in length and beneficial to coverage. Results in Section \ref{sec_eva_cov} have similar conclusions that \textsc{Instiller} has more coverage than DiFuzzRTL.

\subsection{Real-world Vulnerabilities}

To prove the ability to detect real-world bugs of \textsc{Instiller}, we conduct the following evaluation. We collect 12 real-world bugs from the GitHub issue pages of the tested CPUs \cite{boom, mor1kx, or1200, rocket}. As shown in Table \ref{table_bug_real}, the target CPU, bug ID, description, and reproductivity of \textsc{Instiller} are listed. In total, \textsc{Instiller} can reproduce 8 out of the 12 real-world bugs. All of the detected bugs are produced by the bug-reporting mechanism of \textsc{Instiller}, and then they are confirmed by practitioners. This result demonstrates that \textsc{Instiller} has the ability to detect real-world vulnerabilities.

\begin{table}[htbp]
	\setlength{\abovecaptionskip}{-0.1cm}
	\caption{Information of real-world bugs and the detection ability of \textsc{Instiller}}
	\label{table_bug_real}
	\centering
	\resizebox{0.9\columnwidth}{!}{
		\begin{tabular}{cccc}
			\bottomrule
			\bfseries Targets & \textbf{ID} & \bfseries Description & \textbf{Reproducible}\\
			\toprule
			Boom & V1 & \makecell[l]{Instruction count is inaccurate when minstret\\is written by software.} & \CheckmarkBold\\
			Boom & V2 & \makecell[l]{Static rounding is ignored for fdiv.s and fsqrt.s.} & \CheckmarkBold \\
			Boom & V3 & \makecell[l]{Floating point instruction which has invalid rm\\field does not raise exception.} & \CheckmarkBold \\
			Boom & V4 & \makecell[l]{FS bits in mstatus register is set after fle.d\\instruction.} & \XSolidBrush\\
			mor1kx & V5 & \makecell[l]{Incorrect implementation of the carry flag\\generation.} & \XSolidBrush \\
			mor1kx & V6 & \makecell[l]{Missing access checking for privileged register.} & \CheckmarkBold \\
			mor1kx & V7 & \makecell[l]{eear register not saving instruction virtual\\address when illegal instruction exception.} & \CheckmarkBold \\
			mor1kx & V8 & \makecell[l]{l.fl1, l.ff1 instruction decoding fails.} & \CheckmarkBold \\
			or1200 & V9 & \makecell[l]{Incorrect forwarding logic for the GPR0.} & \XSolidBrush \\
			or1200 & V10 & \makecell[l]{Incomplete update logic of overflow bit formsb\\\& mac instructions.} & \CheckmarkBold \\
			or1200 & V11 & \makecell[l]{Incorrect generation of overflow flag.} & \CheckmarkBold \\
			Rocket & V12 & \makecell[l]{EBREAK does not	increase instruction count.} & \XSolidBrush \\
			\bottomrule
		\end{tabular}
	}
\end{table}

\section{Discussion}

\subsection{Golden Reference Models (GRM)}

\textsc{Instiller} and other hardware fuzzers \cite{difuzzrtl, kande2022thehuzz} rely on GRMs to detect hardware bugs. The validation of many commercial CPUs largely depends on the availability of GRMs. There are many industrial large-scale simulators using GRMs, such as Intel x86 Archsim, AMD x86 Simnow, and ARM Cortex Neoverse. Thus, the availability of GRM is not a constraint for \textsc{Instiller}. Moreover, the GRM itself is highly unlikely buggy. They have been carefully designed, which are developed with strict version control, and have undergone extensive testing. Therefore, both in availability and validity, using GRMs in \textsc{Instiller} to detect bugs is not a concern.

\subsection{Different Coverage Metrics}

In this paper, we use control register coverage mentioned in \cite{difuzzrtl}. However, there are other coverage metrics in hardware fuzzing. In RFuzz \cite{rfuzz}, mux control coverage is utilized. TheHuzz \cite{kande2022thehuzz} considers multiple coverage metrics, including branch coverage, condition coverage, FSM coverage, etc. Moreover, software coverage is directly used on the translated model of the hardware RTL in \cite{trippel2022fuzzing}. Although coverage metrics of hardware fuzzing are not the research scope of this paper, we still conducted preliminary evaluations on the metrics. Generally speaking, control register coverage performs the best among the metrics in the evaluation. Therefore, we utilize it in this paper.

\subsection{Hyper-parameters}

There are three hyper-parameters in the design of \textsc{Instiller}, including the tunable parameter in Equation \ref{equ_pheromone}, the proportion of basic metrics in Equation \ref{equ_heu}, and the probability of each mutation operator in Equation \ref{equ_mut}. For Equation \ref{equ_pheromone}, the parameter $\rho$ will affect the strength of mutual influence between the ants, which is related to the global search ability and convergence speed of the algorithm. We compare the three recommended configurations in \cite{dorigo2006ant}, including 0.02, 0.1, and 0.5. Table \ref{table_p} shows the comparison of different configurations of $\rho$. The execution speed of these configurations is similar, and 0.5 reaches the maximum in coverage and mismatches. Therefore, we choose 0.5 as the configuration of $\rho$ in Equation \ref{equ_pheromone}.

\begin{table}[htbp]
	\setlength{\abovecaptionskip}{-0.1cm}
	\caption{Evaluation on parameter $\rho$ in Equation \ref{equ_pheromone} about coverage, speed, and mismatches}
	\label{table_p}
	\centering
	\begin{threeparttable}
		\resizebox{0.7\columnwidth}{!}{
			\begin{tabular}{cccccc}
				\bottomrule
				\bfseries $\rho$ & \bfseries 0.02 & \textbf{0.1} & \textbf{0.5} \\
				\toprule
				Coverage & 2265493.18 & 261107.05 & \textbf{279929.15} \\
				Speed & 0.27 & 0.26 & \textbf{0.27} \\
				Mismatches & 1501.8 & 1511.6 & \textbf{1571.9} \\
				\bottomrule
			\end{tabular}
		}
	\end{threeparttable}
\end{table}

Moreover, we conduct experiments to discuss the parameter $w$ in Equation \ref{equ_heu}. This parameter controls the balance between the basic metrics and the RTL metrics. The basic metrics are related to coverage and execution speed, and the RTL metrics are related to CPU bugs.

\begin{table}[htbp]
	\setlength{\abovecaptionskip}{-0.1cm}
	\caption{Evaluation on parameter $w$ in Equation \ref{equ_heu} about coverage, speed, and mismatches}
	\label{table_w}
	\centering
	\begin{threeparttable}
		\resizebox{0.9\columnwidth}{!}{
			\begin{tabular}{cccccc}
				\bottomrule
				\bfseries $w$ & \bfseries 0.1 & \textbf{0.5} & \textbf{1.0} & \textbf{2.0} & \textbf{10.0}\\
				\toprule
				Coverage & 254329.35 & 257754.12 & 268953.90 & 279929.15 & 281876.22\\
				Speed & 0.24 & 0.25 & 0.25 & 0.27 & 0.31 \\
				Mismatches & 1655.9 & 1641.5 & 1590.1 & 1571.9 & 1105.8 \\
				\bottomrule
			\end{tabular}
		}
	\end{threeparttable}
\end{table}

\vspace{-0.4cm}

Table \ref{table_w} shows how $w$ in Equation \ref{equ_heu} influences the results of coverage, execution speed, and the number of mismatches. As $w$ increases, coverage and speed also increase, and the number of mismatches decreases. We select 2.0 as the configuration of $w$. This value keeps the balance between coverage, speed, and the number of mismatches.

Besides, we conduct experiments to discuss the length parameter $l$ that determines the choices of the mutation operators in Equation \ref{equ_mut}. If the length of the current fuzzing exceeds $l$, the ``deletion" operator will be chosen. Otherwise, ``insertion" is selected. Table \ref{table_l} shows the comparison of three different configurations of $l$. This parameter can affect the input length of fuzzing. We choose 400 as the configuration in our implementation, as it keeps a balance between coverage, speed, and mismatches, compared with other configurations.

\begin{table}[htbp]
	\setlength{\abovecaptionskip}{-0.1cm}
	\caption{Evaluation on parameter $l$ in Equation \ref{equ_mut} about coverage, speed, and mismatches}
	\label{table_l}
	\centering
	\begin{threeparttable}
		\resizebox{0.7\columnwidth}{!}{
			\begin{tabular}{cccccc}
				\bottomrule
				\bfseries $l$ & \bfseries 100 & \textbf{400} & \textbf{1000} \\
				\toprule
				Coverage & 235617.60 & {279929.15} & 251265.47 \\
				Speed & 0.33 & {0.27} & 0.20 \\
				Mismatches & 1211.7 & {1571.9} & 1562.3\\
				\bottomrule
			\end{tabular}
		}
	\end{threeparttable}
\end{table}

\vspace{-0.4cm}

\subsection{Power Schedule}

Power schedule is not mentioned in this paper, which is commonly seen in fuzzing tools \cite{afl, afl++}. If we treat fuzzing as an optimization problem, there are many methods to reach the desired optimum, including seed selection and mutation, coverage metrics, power schedule, etc. The key mechanism of this paper is the VACO algorithm, which shortens the input length, increases coverage, and keeps the input sequence efficient in fuzzing the hardware. As the evaluation shows, VACO accomplishes these tasks with its internal design. Besides, there are other approaches in INSTILLER that make it more realistic in hardware fuzzing, including hardware-based seed selection and mutation. The above components contribute to the ``efficient and realistic" fuzzing of INSTILLER. However, we are not claiming that the power schedule is not important in fuzzing. It is not the research focus of INSTILLER, so it is not included in our paper. Besides, we are preparing to utilize the power schedule in hardware fuzzing and leave this as future work. 

\subsection{Remaining Challenges and Future Improvements}

\textbf{Remaining challenges. }The simulation of RTL execution is relatively slow compared with binary fuzzing, e.g., AFL. Binary fuzzing can reach the speed of thousands of executions per second, and RTL fuzzing can only reach one execution per several seconds. It is still a challenge to speed up the execution speed of RTL simulation.

In software instrumentation, LLVM and Clang greatly reduce the workload of practitioners. However, the instrumentation of CPU RTL requires FIRRTL. It is a time-consuming process to instrument the code.

\textbf{Future improvements. }In the future, multiple ISAs can be added to \textsc{Instiller}, such as ARM ad X86. It would be more attractive for manufacturers to invest in fuzzing these commercial ISAs. However, unlike RISC-V, it is relatively difficult for researchers to get access to these resources to conduct research.

We also plan to propose a new coverage mechanism, aiming to solve different types of bugs in fuzzing the CPU. For example, for bugs related to side channels, such as Meltdown, the coverage mechanism should focus on the branch predictions in the RTL.

%

\section{Related Work}

\subsection{RTL Verification and Testing}
RTL testing is a popular field in research, even in capture-the-flag competitions \cite{chen2022trusting}. RFuzz utilizes mux coverage \cite{rfuzz} to fuzz CPU. DiFuzzRTL applies differential fuzz testing \cite{difuzzrtl}, which is a pioneer in this field. TheHuzz uses different coverage metrics \cite{kande2022thehuzz} and conducts experiments to show their performance. SpecDoctor is an automated RTL fuzzer to find transient execution vulnerabilities \cite{hur2022specdoctor}. GenFuzz is a GPU-accelerated hardware fuzzer with a genetic algorithm and multiple inputs \cite{lin2023genfuzz}. Cascade uses asymmetric ISA pre-simulation to construct RISC-V programs \cite{soltcascade}. \textsc{Instiller} is different from them, and we aim to shorten input instruction length and increase coverage.

HyperFuzzing converts hardware into software \cite{muduli2020hyperfuzzing}, which is a different approach to fuzz RTL. Hardware Fuzzing Pipeline also translates hardware to software to fuzz CPU \cite{trippel2022fuzzing}. Our tool does not need to translate hardware to software. These two tools are different from \textsc{Instiller}.

\subsection{Coverage-guided Grey-box Fuzzing}

CGF becomes prevalent since the release of AFL \cite{afl}. It discovers numerous bugs with its well-designed mechanism. The key to CGF is the coverage feedback to the fuzzer. AFLFast \cite{bohme2016coverage} is another milestone of CGF, which improves the power schedule (the number of executions on a seed). MOPT \cite{mopt} automatically selects mutation operators using Particle Swarm Optimization (PSO), which is a promising direction in fuzzing research. CollAFL proposes a coverage-sensitive fuzzing solution \cite{gan2018collafl}, which uses a finely designed coverage metric to effectively avoid path collisions in fuzzing. ovAFLow \cite{zhang2022ovaflow} deals with the problem of taint input bytes. It utilizes a lightweight fuzzing-based taint inference to guide the mutation strategy. \textsc{Instiller} is different from them, which utilizes the idea of CGF combined with differential testing to detect CPU bugs. More importantly, the goal of \textsc{Instiller} is to distill input instructions for more efficient hardware fuzzing, and it is different from the above fuzzers.

\subsection{Optimization in Fuzzing}

As a classical optimization technique, ACO is used in fuzzing. ACOFuzz uses ACO to allocate energy \cite{wu2022acofuzz}, which controls the power schedule of fuzzing. AFL-ant proposes a seed screening technique based on ACO \cite{sun2020greybox}. This technique chooses the best seed with ACO. RGF concentrates on new code with an ACO-based power schedule \cite{zhu2021regression}. Therefore, it is a directed fuzzer. In \textsc{Instiller}, based on the characteristics of RTL fuzzing, we propose a variant of ACO to distill input instructions. Our VACO is different from the classic ACO in the above-related work. \textsc{Instiller} and these ACO-based fuzzers focus on different aspects of the fuzzing process.

Other optimization techniques can also be adopted in fuzzing. For example, EcoFuzz \cite{yue2020ecofuzz} uses a variant of multi-arm bandit (MAB) to assign energy in fuzzing. The power schedule is altered according to the fuzzing status. MobFuzz \cite{zhang2022mobfuzz} solves the problem of multi-objective optimization with a method called multi-player MAB. The optimization technique in \textsc{Instiller} is ACO, which is different from the MAB algorithm, and we make modifications to it according to the situation of RTL fuzzing.

\subsection{Seed Selection and Mutation}

Seed selection and mutation strategies can improve fuzzing efficiency. TortoiseFuzz uses three metrics to select and prioritize seeds \cite{wang2020not}. FairFuzz selects seeds that are chosen less frequently \cite{fairfuzz}. MemLock uses the amount of memory consumption to detect memory consumption bugs \cite{memlock}. FuzzFactory proposes waypoints to guide seed selection and improve fuzzing \cite{fuzzfactory}. Waypoints can be memory consumption, algorithmic complexity, etc. MoonShine \cite{pailoor2018moonshine} optimizes OS fuzzer seed selection with trace distillation. It can distill seeds from system call traces of real-world programs while preserving the dependencies across the system calls. \cite {herrera2021seed} compares six seed selection approaches, concluding that fuzzing highly relies on the initial seed corpus. However, none of them is related to the heuristics of hardware, e.g., the number of jump instructions. \textsc{Instiller} uses hardware-related strategies, which improves the RTL fuzzing efficiency.

\section{Conclusion}
In this paper, we conclude three challenges in RTL fuzzing in previous work, which include increasing input instruction length, no realistic interruption or exception, and no hardware-related seed selection and mutation. To solve these challenges, we propose input instruction relationship extraction, together with input instruction distillation based on a variant of ant colony optimization. Moreover, we enable our fuzzer to include multiple interruptions and exceptions to cover more RTL states. We also propose hardware-related seed selection and mutation strategies to improve the fuzzing performance. In addition, we implement a prototype \textsc{Instiller} and conduct extensive experiments against state-of-the-art fuzzing work. The results show our tool outperforms previous work in coverage, input instruction length, and vulnerability discovery, which demonstrate the effectiveness of our proposed techniques.

\section*{Acknowledgment}
Thanks to the reviewers' valuable comments and helpful suggestions. This work is supported by the Natural Science Foundation of China (61902412, 61902416), the Research Project of National University of Defense Technology (ZK20-17, ZK20-09, and ZK23-14), the Natural Science Foundation of Hunan Province of China (2021JJ40692).

\bibliographystyle{unsrt}
\bibliography{ref}

\vspace{-1cm}

\begin{IEEEbiography}
	[{\includegraphics[width=0.5\columnwidth, clip, keepaspectratio]{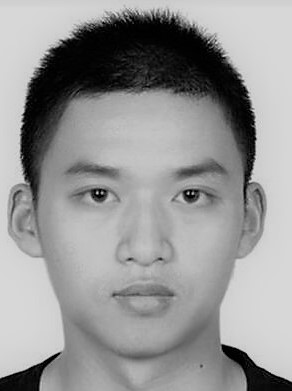}}]
	{\bf Gen Zhang} received his Ph.D. degree in computer science and technology in 2022 from NUDT. His research interests include fuzzing and testing.
\end{IEEEbiography}

\vspace{-2cm}

\begin{IEEEbiography}
	[{\includegraphics[width=0.5\columnwidth, clip, keepaspectratio]{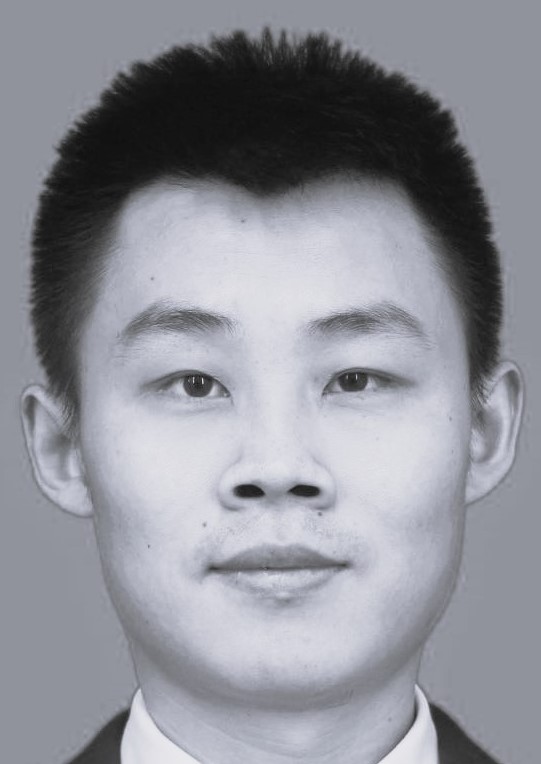}}]
	{\bf Pengfei Wang} received his B.S., M.S., and Ph.D. degrees in computer science and technology, in 2011, 2013, and 2018, respectively, from NUDT. His research interests include operating system and software testing.
\end{IEEEbiography}

\vspace{-2cm}

\begin{IEEEbiography}
	[{\includegraphics[width=0.5\columnwidth, clip, keepaspectratio]{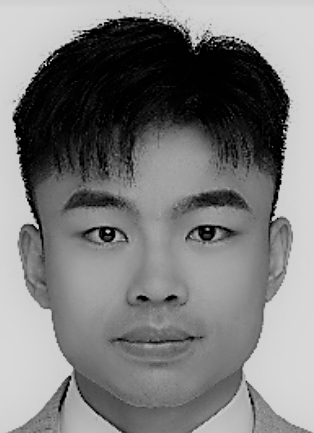}}]
	{\bf Tai Yue} received his B.S. and M.S. degrees in computer science and technology, in 2017 and 2019 from Nanjing University and NUDT. His research interests include fuzzing and testing.
\end{IEEEbiography}

\vspace{-2cm}

\begin{IEEEbiography}
	[{\includegraphics[width=0.5\columnwidth, clip, keepaspectratio]{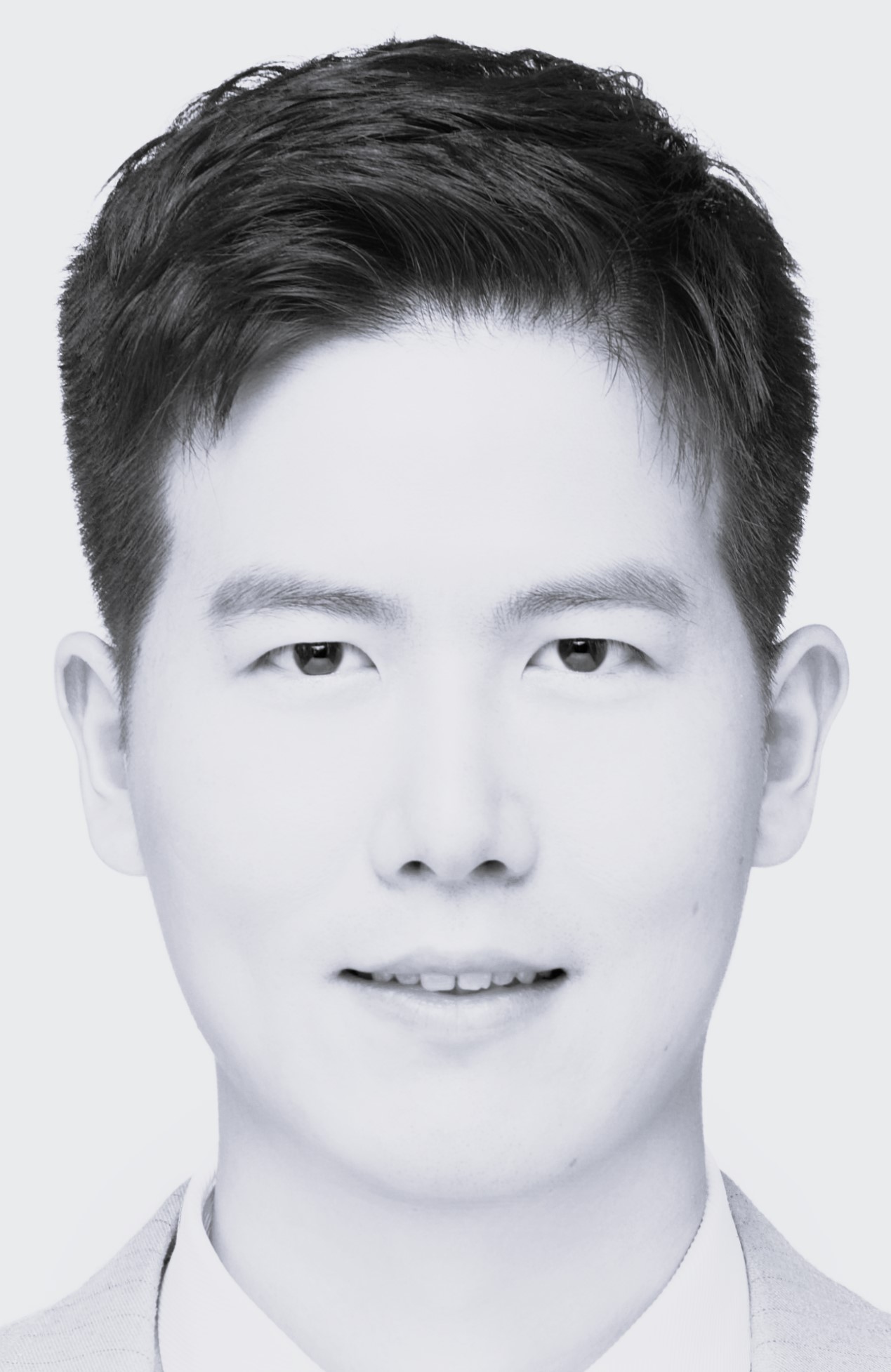}}]
	{\bf Danjun Liu} received his Ph.D. degree in computer science and technology in 2023 from NUDT. He is now an assistant professor in NUDT. His research interests include operating systems and parallel computing.
\end{IEEEbiography}

\vspace{-2cm}

\begin{IEEEbiography}
	[{\includegraphics[width=0.5\columnwidth, clip, keepaspectratio]{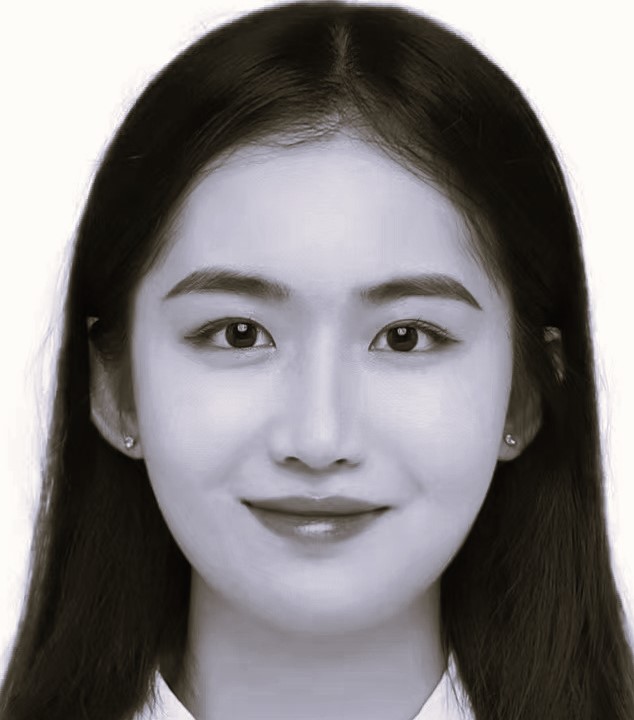}}]
	{\bf Yubei Guo} received her master degree in computer science and technology in 2020, from Hunan University. Her research interests include security.
\end{IEEEbiography}

\vspace{-2cm}

\begin{IEEEbiography}
	[{\includegraphics[width=0.5\columnwidth, clip, keepaspectratio]{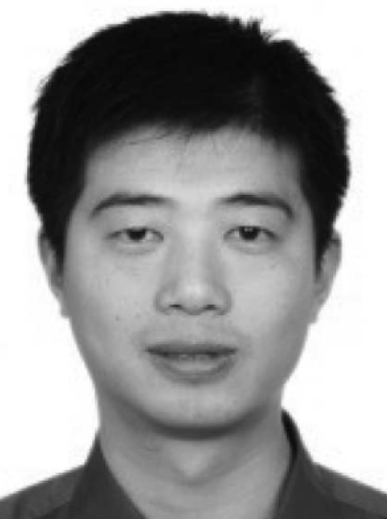}}]
	{\bf Kai Lu} received his B.S. degree and Ph.D. degree in 1995 and 1999, respectively, both in computer science and technology from NUDT. He is now a professor in NUDT. His research interests include operating systems, parallel computing, and security.
\end{IEEEbiography}

\end{document}